\documentclass[prx,twocolumn,amsmath,amssymb,superscriptaddress,bibliography,aps]{revtex4-2} 

\usepackage{color,times}
\usepackage{bm}
\usepackage{dcolumn}
\usepackage{multirow}
\usepackage{amssymb,amsfonts,amsmath,graphicx}
\usepackage{epsfig}
\usepackage{xcolor}
\usepackage{SIunits}
\usepackage{braket}
\usepackage{epstopdf}
\usepackage{stmaryrd}
\usepackage{amsmath}
\usepackage[utf8]{inputenc}
\usepackage[english]{babel}
\usepackage[colorlinks=true, breaklinks=true, linkcolor=darkblue, citecolor=darkblue, urlcolor=darkblue]{hyperref}
\usepackage{relsize}
\usepackage{ulem}
\usepackage{verbatim}

\definecolor{darkblue}{rgb}{0, 0, 0.8}
\definecolor{darkgreen}{rgb}{0, 0.5, 0}
\definecolor{violet_custom}{rgb}{0.8549019607843137, 0.4392156862745098, 0.8392156862745098}
\definecolor{blue_custom}{rgb}{0.       , 0.       , 0.4728164}
\definecolor{gray_custom}{rgb}{0.5019607843137255, 0.5019607843137255, 0.5019607843137255}
\definecolor{orange_custom}{rgb}{0.93, 0.47, 0.2}

\graphicspath{{Figures_vfinal//}}

\renewcommand{\selectlanguage}[1]{} 

\addtocounter{equation}{0}

\begin{document}
	
\title{Tomonaga-Luttinger Liquid Behavior in a Rydberg-encoded Spin Chain}

\author{Gabriel Emperauger}
\thanks{GE, MQ, CC, FC, SB and MB contributed equally to this work.}
\affiliation{Université Paris-Saclay, Institut d'Optique Graduate School,\\
CNRS, Laboratoire Charles Fabry, 91127 Palaiseau Cedex, France}

\author{Mu Qiao}
\thanks{GE, MQ, CC, FC, SB and MB contributed equally to this work.}
\affiliation{Université Paris-Saclay, Institut d'Optique Graduate School,\\
CNRS, Laboratoire Charles Fabry, 91127 Palaiseau Cedex, France}

\author{Cheng Chen}
\thanks{GE, MQ, CC, FC, SB and MB contributed equally to this work.}
\affiliation{Université Paris-Saclay, Institut d'Optique Graduate School,\\
CNRS, Laboratoire Charles Fabry, 91127 Palaiseau Cedex, France}
\affiliation{Institute of Physics, Chinese Academy of Sciences, Beijing 100190, China}

\author{Filippo Caleca}
\thanks{GE, MQ, CC, FC, SB and MB contributed equally to this work.}
\affiliation{Univ Lyon, Ens de Lyon, CNRS, Laboratoire de Physique, F-69342 Lyon, France}

\author{Saverio Bocini}
\thanks{GE, MQ, CC, FC, SB and MB contributed equally to this work.}
\affiliation{Univ Lyon, Ens de Lyon, CNRS, Laboratoire de Physique, F-69342 Lyon, France}

\author{Marcus Bintz}
\thanks{GE, MQ, CC, FC, SB and MB contributed equally to this work.}
\affiliation{Department of Physics, Harvard University, Cambridge, Massachusetts 02138 USA}

\author{Guillaume Bornet}
\affiliation{Université Paris-Saclay, Institut d'Optique Graduate School,\\
CNRS, Laboratoire Charles Fabry, 91127 Palaiseau Cedex, France}

\author{Romain Martin}
\affiliation{Université Paris-Saclay, Institut d'Optique Graduate School,\\
CNRS, Laboratoire Charles Fabry, 91127 Palaiseau Cedex, France}

\author{Bastien G\'ely}
\affiliation{Université Paris-Saclay, Institut d'Optique Graduate School,\\
CNRS, Laboratoire Charles Fabry, 91127 Palaiseau Cedex, France}

\author{Lukas Klein}
\affiliation{Université Paris-Saclay, Institut d'Optique Graduate School,\\
CNRS, Laboratoire Charles Fabry, 91127 Palaiseau Cedex, France}

\author{Daniel Barredo}
\affiliation{Université Paris-Saclay, Institut d'Optique Graduate School,\\
CNRS, Laboratoire Charles Fabry, 91127 Palaiseau Cedex, France}
\affiliation{Nanomaterials and Nanotechnology Research Center (CINN-CSIC), 
Universidad de Oviedo (UO), Principado de Asturias, 33940 El Entrego, Spain}

\author{Shubhayu~Chatterjee}
\affiliation{Department of Physics, Carnegie Mellon University, Pittsburgh, PA 15213, USA}

\author{Norman Yao}
\affiliation{Department of Physics, Harvard University, Cambridge, Massachusetts 02138 USA}

\author{Fabio~Mezzacapo}
\affiliation{Univ Lyon, Ens de Lyon, CNRS, Laboratoire de Physique, F-69342 Lyon, France}

\author{Thierry~Lahaye}
\affiliation{Université Paris-Saclay, Institut d'Optique Graduate School,\\
		CNRS, Laboratoire Charles Fabry, 91127 Palaiseau Cedex, France}	
		
\author{Tommaso~Roscilde}
\affiliation{Univ Lyon, Ens de Lyon, CNRS, Laboratoire de Physique, F-69342 Lyon, France}

\author{Antoine~Browaeys}
\email{antoine.browaeys@institutoptique.fr}
\affiliation{Université Paris-Saclay, Institut d'Optique Graduate School,\\
		CNRS, Laboratoire Charles Fabry, 91127 Palaiseau Cedex, France}

\begin{abstract}
Quantum fluctuations can disrupt long-range order in one-dimensional systems, and replace 
it with the universal paradigm of the Tomonaga-Luttinger liquid (TLL), 
a critical phase of matter characterized by power-law decaying correlations and linearly dispersing excitations.
Using a Rydberg quantum simulator, we study how TLL physics manifests in the 
low-energy properties of a spin chain, interacting under either the ferromagnetic 
or the antiferromagnetic dipolar XY Hamiltonian.
Following quasi-adiabatic preparation, we directly observe the power-law decay 
of spin-spin correlations in real-space, allowing us to extract the Luttinger parameter. 
In the presence of an impurity, the chain exhibits tunable Friedel oscillations of the local magnetization. 
Moreover, by utilizing a quantum quench, we directly probe the propagation of correlations, 
which exhibit a light-cone structure related to the linear sound mode of the underlying TLL. 
Our measurements demonstrate the influence of the long-range dipolar interactions,  
renormalizing the parameters of TLL with respect to the case of nearest-neighbor interactions. 
Finally, comparison to numerical simulations exposes the high sensitivity of TLLs to doping and finite-size effects. 
\end{abstract}

\date{\today}

\maketitle

\section{Introduction}\label{Sec:Introduction}

Quantum physics in one dimension (1D) can show radically new phenomena compared to higher dimensions. 
This comes from the enhanced role of quantum fluctuations, that generically
suppress classical long-range order~\cite{Giamarchi_2004}. 
The low-energy properties of gapless 1D systems with short-range interactions are described 
by a universal harmonic quantum field theory of free massless bosons,
the Tomonaga-Luttinger liquid (TLL)~\cite{Bethe_1931, Tomonaga_1950, Luttinger_1960, Haldane_1981, Giamarchi_2004}.
Remarkably, TLL theory predicts that all long-wavelength properties  of the system can be related 
to the knowledge of only two numbers: the dimensionless stiffness (Luttinger parameter), $K$, and the sound velocity, $u$.  
This theory not only describes the low-energy physics of  bosonic 
systems \cite{Cazalilla_2011} but also of spin chains~\cite{Giamarchi_2004} 
and interacting fermions in 1D~\cite{Tomonaga_1950, Luttinger_1960, Giamarchi_2004}.

The universality associated with TLLs has rendered them crucial testbeds for both theory and experiment.
TLL physics has been widely explored in experiments in 
condensed matter 
\cite{Yacoby_1996,  Grayson_1998, Schwartz_1998, Bockrath_1999, Auslaender_2002,
	Jerome_2004, Lee_2004, Lake_2005, Klanjsek_2008, Li_2024} and in cold 
atoms~\cite{Paredes_2004, Kinoshita_2004, Fabbri_2015, Hofferberth_2008, 
	Yang_2017, Yang_2018, Pagano_2014, Hilker_2017, Seneratne_2022,Cavazos-Cavazos_2023}.
In the context of quantum spin systems -- which is the focus of this work -- TLL physics has been extensively probed in
spin-chain and spin-ladder materials \cite{zheludevQuantumCriticalDynamics2020}. 
In these systems, residual couplings between the chains or ladders inevitably lead to long-range order at low temperatures, 
preventing the observation of ground-state TLL physics. The physics of TLLs can still be probed at sufficiently high temperature, 
where the dynamical response functions at equilibrium can reveal the value of the $u$ and $K$ 
parameters \cite{Klanjsek_2008, zheludevQuantumCriticalDynamics2020}.
However, some important features of TLLs, such as the power-law nature of correlations upon 
approaching the ground state, or their non-equilibrium behavior, remain elusive.

Over the past decades, experimental progress has made it possible to create 
and manipulate one-dimensional systems with 
single-particle control~\cite{Jurcevic_2014, Richerme_2014, Hilker_2017, Morvan_2022, Fang_2024, Kim_2024},  
enabling precise tests of our understanding of 1D physics. 
In this work, we explore the low-energy, TLL physics of a 1D ring 
of {\it dipolar}-interacting Rydberg atoms. 
The controllability and single-particle resolution of our experiments enable direct 
measurements of real-space correlations, local susceptibilities, and real-time dynamics.
Our experiment realizes a power-law interacting spin-1/2 XY chain, 
with either ferromagnetic (FM) or antiferromagnetic (AFM) couplings.
Much like the paradigmatic XY chain with nearest-neighbor (NN) interactions -- which is dual to free fermions with $K=1$ \cite{Lieb_1961} -- 
our system is in both cases predicted realize a TLL at low energies ~\cite{Maghrebi_2017,Schneider_2022,Gupta_2023,Lee_2023}.
However, the extended dipolar interactions either reinforce (FM) or frustrate (AFM) each other, modifying $K$ and $u$ from their NN values.
Notably, we find that for the FM the stiffness is enhanced, $K_{\rm FM} >1$, similarly to the behavior of 
attractively interacting fermions; while for the AFM, $K_{\rm AFM} <1$,  as with repulsive fermions.

We perform three different experiments, each one probing a different aspect of TLL behavior. 
First, we use local atomic control to adiabatically prepare low-energy states for 
both the FM and the AFM, and then measure the resulting spin-spin correlations in real-space.
In multiple measurement bases, the spatial profiles of these correlations feature a power-law decay, 
from which we directly estimate the Luttinger parameters $K_{\rm{FM}}$ and $K_{\rm{AFM}}$. 
Second, we cut the AFM chain by removing a single atom, and observe Friedel-like oscillations~\cite{Friedel1958} of the 
local magnetization after adiabatic preparation; moreover, we demonstrate the ability to linearly tune 
this oscillation wavevector by adjusting the total magnetization of the chain.
Finally, for both FM and AFM chains, we probe the real-time dynamics upon 
quenching from a low-energy product state, and observe the ballistic propagation of correlations.
For the AFM this provides a precise measurement of the sound velocity $u$.
Interestingly, for the FM, proximity to a continuous symmetry breaking phase 
transition~\cite{Maghrebi_2017,Schneider_2022} yields strong finite size effects 
and we measure an effective velocity significantly smaller than theoretically predicted.

\begin{figure}[h!]
	\centering
	\includegraphics[]{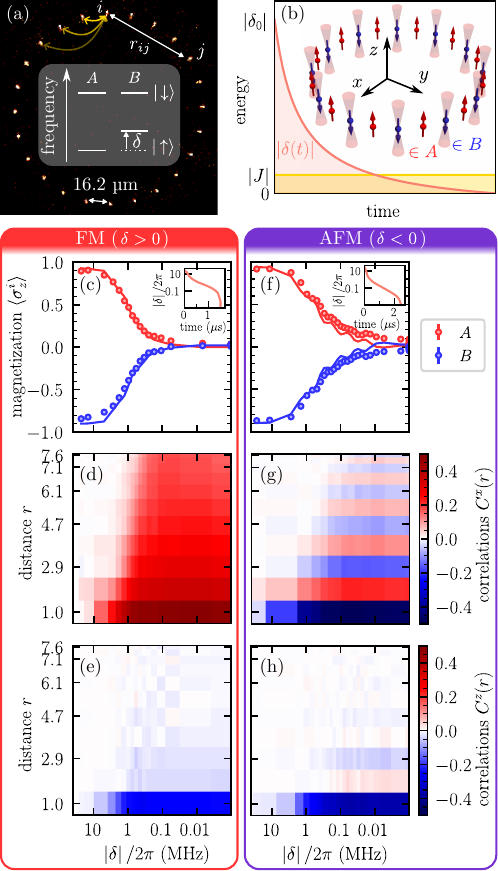}
	\caption{{\bf Quasi-adiabatic preparation of XY FM and AFM.}
		(a)~Geometry of the $N=24$ chain of Rydberg atoms. 
		Yellow arrows between atoms indicate the dipolar XY interaction.
		Inset: definition of the light shift $\delta$ used to shift the energy of the state $\ket{\uparrow}$ 
		on a given set of atoms, defining two sublattices $A$ (non-addressed atoms) and $B$ (addressed atoms).
		(b)~Sketch of the experimental sequence for adiabatic preparation. 
		The addressing light shifts ($\delta(t)$, pink line) are ramped down in the presence 
		of the dipolar XY interactions ($J$, gold line). Inset: representation of the initial state, 
		along with the position of the addressed sites from sublattice~$B$ (pink laser spots).
		(c)~Evolution of the $z$-magnetization per sublattice with the light shift, showing the 
		melting down of the staggered order along $z$. Solid lines are simulations 
		of the dynamics that include experimental errors. 
		Inset: light shift as a function of time.
		(d)~Evolution of the $x$-correlations $C^x (r)$ for all distances $r$, revealing the construction of the FM order.
		(e)~Evolution of the $z$-correlations $C^z (r)$.
		(f,g,h)~ Same as (c,d,e) for the AFM case.
	}
	\label{fig:fig1}
\end{figure}

\section{Experimental system}\label{Sec:System}

Using optical tweezers generated by a spatial light modulator (SLM)~\cite{Nogrette_2014, Schymik_2020}, 
we trap $N=24$ rubidium atoms in a circular geometry with a distance of $16.2\,\mu$m 
between nearest neighbors [Fig.~\ref{fig:fig1}(a)] 
\footnote{Since the diameter of the circle ($124$~$\mu$m) is large compared with 
the nearest-neighbor distance, we do not expect strong deviations from an ideal one-dimensional chain.}.
We encode a pseudo-spin $1/2$ using the two Rydberg states 
$\ket{\uparrow} = \ket{70S_{1/2},m_J=1/2}$ and  $\ket{\downarrow} = \ket{70P_{1/2},m_J=-1/2}$. 
The resonant dipole interaction realizes a dipolar XY spin Hamiltonian:
\begin{equation}\label{eq:Hamiltonian}
	H_{\rm XY} = -{\frac{\hbar J}{2}}
	\sum_{i < j}  \frac{1}{r_{ij}^3}  \left(\sigma^x_i \sigma^x_j + \sigma^y_i \sigma^y_j \right).
\end{equation}
Here, $J \approx 2\pi \times 0.55~$MHz is the nearest-neighbor interaction strength, $r_{ij}$ 
is the distance between atoms $i$ and $j$ (in units of the nearest-neighbor distance), 
and $\sigma_i^{\mu}$ are the Pauli matrices acting on spin $i$ in direction $\mu$. 
This Hamiltonian exhibits a $U(1)$ symmetry, and consequently the total $z$-magnetization 
$M_z = \sum_i \sigma_i^z$ is conserved.

On top of the naturally-occurring interactions, we create an effective
magnetic field with controlled amplitude in space and time~\cite{Chen_2023}. 
To this end, we use an addressing laser at $1014$~nm with a detuning $\Delta$ 
from the atomic transition $70S_{1/2} - 6P_{3/2}$. This induces a light shift $\delta$ 
on the level $\ket{\uparrow}$ of the addressed atoms: 
$\delta\approx \Omega^2 / (4\Delta)$ with $\Omega$ the Rabi frequency of 
the addressing light [inset of Fig.~\ref{fig:fig1}(a)]. A dedicated SLM  
generates addressing spots on the desired atoms, and an acousto-optic modulator 
controls the overall amplitude in time. The addressing Hamiltonian reads 
$H_{\rm Z}(t) =\sum_{i \in B} \hbar \delta(t) (1+\sigma^z_i)/2$, 
where the sublattice $B$ denotes the addressed atoms. 
The sign of $\delta$, fixed during a sequence, is given by that of $\Delta$ which we can chose arbitrarily.

\section{Adiabatic preparation of a closed ring}\label{Sec:GS}

\begin{figure*}
	\centering
	\includegraphics{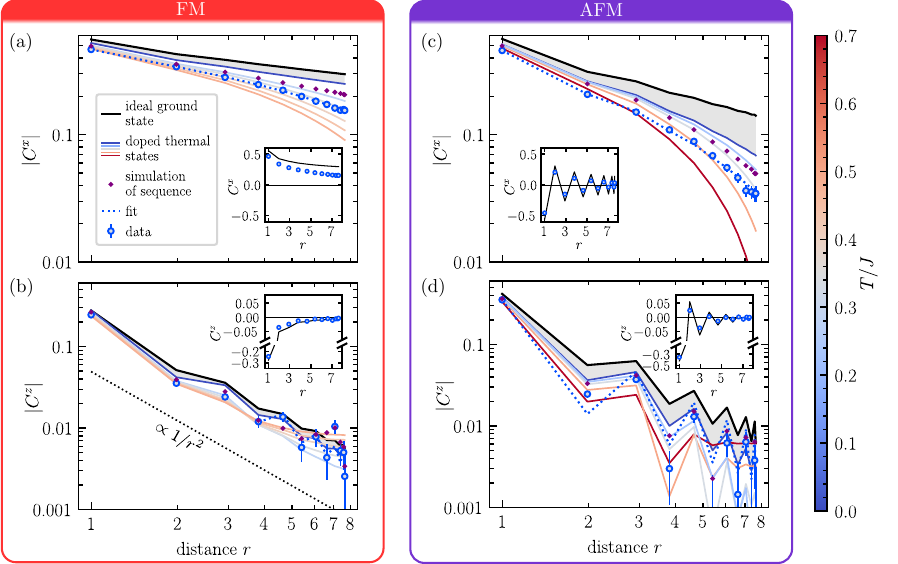}
	\caption{{\bf Spatial profiles of the correlations in the ferromagnetic (FM) and antiferromagnetic (AFM) ground states.}
	The left panels show the FM correlations at $t=1600$~ns along $x$~(a) and along $z$~(b). 
	The right panels display the AFM correlations along $x$ at $t=3000$~ns~(c) and along $z$ at $t=3200$~ns~(d).
	The blue points represent experimental data, while the purple points correspond to simulated 
	sequences with imperfections. The dotted lines indicate fits by the theoretical profiles (see text).
	The solid lines depict simulations of the ideal ground state (in black) and thermal states in the presence 
	of randomly placed holes, with the color of the lines coding for the temperature. 
	For a fair comparison with the data, all simulated correlations are multiplied by a factor of~$0.89$, 
	corresponding to experimentally measured detection errors 
	(see App.~\ref{SM:Experimental_errors}). 
	The doping with holes is $4\%$ in the FM case and $6\%$ in the AFM case, 
	and its effect is highlighted by the grey region.
	}
	\label{fig:fig2}
\end{figure*}

We first consider a closed ring geometry that realizes periodic boundary conditions (PBC). 
This geometry reduces finite-size effects compared to an open-boundary condition (OBC) chain, 
and enables improved statistics of observables by averaging over sites, which are all equivalent.
We aim at preparing  a low-energy state, close to the ground state of the Hamiltonians $\pm H_{\rm XY}$.
To do so, we use a quasi-adiabatic scheme similar to Refs.~\cite{Chen_2023,fengContinuousSymmetryBreaking2023}.
First, we apply large light shifts $\left|\delta_0 \right| \approx 2\pi \times 23$~MHz $\gg \left| J \right|$ 
on a staggered sublattice~$B$ [inset of Fig.~\ref{fig:fig1}(b)]; 
then, we prepare the ground state of the addressing Hamiltonian $H_{\rm Z}$ at half-filling ($M_z = 0$), 
which is the product state $\ket{\psi_0} = \ket{\uparrow \downarrow \cdots \uparrow \downarrow}$ 
[inset of Fig.~\ref{fig:fig1}(b)]; finally, we ramp down the light shifts $\delta(t)$ using an adiabatic  
profile~\cite{Richerme_2013, Fang_2024}, in order to end up in a state close to the ground state of $H_{\rm XY}$. 
The sequence is illustrated in Fig.~\ref{fig:fig1}(b), and is described in more details in 
App.~\ref{SM:method_adiabatic_preparation}. 
The duration and shape of the ramp were experimentally optimized 
to limit the effects of decoherence, while preserving adiabaticity.
The same protocol is used to prepare a state close to the AFM ground state of $-H_{\rm XY}$, 
by  changing the sign of $\delta(t)$: with $\delta<0$, the initial state $\ket{\psi_0}$
is the highest energy state of $H_{\rm Z}$, and an adiabatic preparation connects 
$\ket{\psi_0}$ to the highest-energy state of $H_{\rm XY}$, i.e. the ground state of $-H_{\rm XY}$. 

During the adiabatic ramp, we monitor the dynamics of the spins. 
Figures~\ref{fig:fig1}(c) and (f) show the $z$-magnetization of each sublattice, respectively in the FM and AFM cases.
At the beginning of the preparation, the $z$-magnetizations of sublattices $A$ and $B$ are opposite, 
reflecting the staggered spin pattern of the initial state $\ket{\psi_0}$. 
As we ramp down the light shifts, they merge to zero,  signaling the meltdown of the initial pattern 
into a translation-invariant state in the $xy$ plane.
To further characterize the FM and AFM states prepared during the ramp, we measure
the connected correlations between the $\sigma^\mu$ spin components $(\mu \in \{x,z\})$  for all spin pairs $(i,j)$, defined as 
$C_{i,j}^{\mu} = 
\langle \sigma^{\mu}_i \sigma^{\mu}_j \rangle - \langle \sigma^{\mu}_i \rangle \langle \sigma^{\mu}_j \rangle$. 
Figures~\ref{fig:fig1}(d,e,g,h) present the correlations $C^{\mu}(r)$ averaged over pairs 
separated by the same chord distance~$r$~\footnote{We use the chord distance $r_{ij}$ instead 
of the perimeter distance $|i-j|$ to avoid biases due to the periodic boundary conditions. 
The link between the two is  $r_{ij} = {N \over \pi} \sin(\pi |i-j|/ N)$. 
A theoretical justification relies on conformal field theory, 
see for instance~\cite{Di_Francesco_2011,Slagle_2021}.}.
As expected, we observe the progressive 
buildup of FM or AFM correlations along $x$ as $\delta(t)$ is ramped down to 0 [Fig.~\ref{fig:fig1}(e,h)]. 
Along the $z$ axis, we observe weak negative correlations in the FM case and weak
staggered magnetization in the AFM case [Fig.~\ref{fig:fig1}(d,f)].

\section{Critical correlations}

We now examine in detail the correlations of the final state, shown in Fig.~\ref{fig:fig2}.
TLL theory predicts that these correlations are scale-invariant, indicative of a quantum critical state. 
More specifically, $C^x$ and $C^z$ should decay with distance as sums of power-laws, 
whose exponents are universally determined by the single Luttinger parameter, $K$.
We first focus on the FM case in the $x$ basis, for which the dominant power-laws 
(i.e. at asymptotic distances $r\gg 1$) are expected to be~\cite{Giamarchi_2004}:
\begin{equation}\label{eq:Cx_FM}
C^x_{\rm FM}(r) \approx
A \left( \frac{1}{r} \right)^{\mathlarger{\frac{1}{2 K}}}
+ B \left( -1 \right)^{d(r)}
\left( \frac{1}{r} \right)^{2K + \mathlarger{\frac{1}{2 K}}}
\end{equation}
where $A$ and $B$ are non-universal amplitudes, and 
$d(r) \equiv \frac{N}{\pi} \arcsin\left(\frac{\pi r}{N}\right)$ is the perimeter distance in units of the lattice spacing. 
Numerical calculations (see App.\ref{SubSM:Simulation_of_the_quench_experiment}) 
predict $K_{\text{FM}} = 1.85(1)$ for our $N=24$ atom ring. 

The FM experimental data feature a power-law decay of the $x$-correlations up to $\sim 5$ sites 
[blue points on Fig.~\ref{fig:fig2}(a)]. 
At larger distances, the correlations decay faster than a power-law;  
we attribute this deviation to preparation errors (see below, and App.~\ref{SM:Experimental_errors}). 
To account for this deviation, we follow the approach of~\cite{Gori_2016, Fang_2024} and use a 
modified fit function for the data: $\tilde{C}^x_{\rm FM}(r) = C^x_{\rm FM}(r) e^{-r/\xi}$,  
where $\xi$ is an empirical correlation length. 
We obtain $K_{\rm FM} = 1.6(4)$, already close to the theoretical value, 
and a correlation length $\xi = 15(4)$~sites. 

Next, we examine the FM $z$-correlations, whose theoretical behavior at long distance takes the form,
\begin{equation}\label{eq:Cz_FM}
	C^z(r) \approx -\frac{2K}{\pi^2} \left( \frac{1}{r} \right)^{2} + D \left( -1 \right)^{d(r)} \left( \frac{1}{r} \right)^{2K}~.
\end{equation}
The structure of the first term, with an integer exponent and $K$ appearing as a simple 
prefactor, encodes the special role of $\sigma^z$ as the local density of a conserved quantity, $M_z$.
By contrast, the second term reflects the presence of emergent gapless fluctuations at wavevector $k=\pi$, 
which, similar to the $x$ correlations, has a non-universal amplitude, $D$, and a $K$-dependent exponent.
In the measured correlations, the $1/r^2$ term dominates, as highlighted by the black dotted line in Fig.~\ref{fig:fig2}(b).
We find that Eq.~(\ref{eq:Cz_FM}) fits the experimental data well, without the need for an exponential correction, 
but we face two challenges in using it to determine $K$.
First, for the FM, the staggered part has a small amplitude and a rapid decay ($2K\approx 3.6$), 
which makes fits of that exponent unstable even in ideal numerical data. 
Second, the uniform prefactor gives a reliable estimate of $K$ in theoretical calculations 
but experimental imperfections and read out errors reduce the overall magnitude 
of measured correlations in a non-universal way.
Accounting for readout errors (see App.\ref{SM:fitting_procedure}), 
our experimental fit to Eq.~(\ref{eq:Cz_FM}) yields $K_{\mathrm{FM}}\approx1.4(1)$.

Proceeding now to the AFM, we expect the same critical correlation structure, 
except with a global staggered sign $(-1)^{d(r)}$ multiplying the $x$-correlations, 
and a different Luttinger parameter, which we estimate as $K_{\rm{AFM}} = 0.85$ 
from our simulation of the ground state of a $N=24$ chain.
The measured $z$-correlations are fitted reasonably well by Eq.~(\ref{eq:Cz_FM}), 
and we obtain  $K_{\mathrm{AFM}} \approx 0.90(1)$ [Fig.~\ref{fig:fig2}(d)]. 
However, the $x$-correlations again require the introduction of an exponential decay, 
with a short correlation length of $\xi=5(1)$ sites [Fig.~\ref{fig:fig2}(c)]; 
this leads to larger uncertainty in our estimate of the Luttinger parameter, $K_{\rm{AFM}}\approx 1.0(3)$. 
To check the robustness of the analysis, we applied the same 
preparation protocol for different system sizes (ranging from $N=16$ to $N=28$) 
and obtained similar values of $K$ and $\xi$.
The small value of $\xi$ indicates that the AFM is particularly sensitive 
to experimental imperfections.

Motivated by this observation, we carried out numerical simulations of the state 
preparation procedure via matrix-product-state (MPS) approaches,
including most known experimental imperfections (see App.\ref{SM:Numerical_methods} and \ref{SM:Luttinger}). 
The results of these calculations (Fig. 2, purple diamonds) quantitatively 
reproduce much of the deviation between the measurements and the ideal ground state.
Among the included errors, we find that a finite density $p$ of holes is especially important,  
i.e. atoms lying outside the Rydberg manifold due to either a failed excitation or spontaneous decay. 
We estimate $p=0.04$ for the FM at the end of the ramp, while $p=0.06$ for the AFM due to a longer adiabatic preparation. 
Owing to the dipolar interactions, sites separated by a hole remain coupled with a reduced strength $J/8$.
However, each hole leads to a slip of the sublattice structure, which causes snapshot-averaged measurements 
of staggered part of the correlations to decay over a perimeter distance $\xi_p = 1/|\ln(1-2p)|$  \cite{Bocini_2024}.  
This disordered-readout effect is most significant in the AFM $x$-correlations, where the staggered part dominates; 
the preceding formula for $\xi_p$ corresponds to a chord distance of 6.5, 
which quantitatively accounts for the observed exponential decay. 
Additionally, when fitting the simulated FM correlations with holes by Eqs.\,(\ref{eq:Cx_FM},\ref{eq:Cz_FM}), 
we obtain  $K_{\rm FM}= 1.55(1)$ from the $C^x$ correlations 
and $K_{\rm FM} = 1.44(1)$ from the $C^z$ correlations, in good agreement with the 
experimental values. 
For the AFM, we expect that this hole density ultimately destabilizes the TLL phase 
in the thermodynamic limit (see App.\ref{subSM:chains_with_holes}), although TLL-like behavior may still emerge at short distance.
Finally, the other imperfections, such as non-adiabaticity, effectively raise the energy of the final prepared state:  
comparing to equilibrium calculations that include holes 
(based on both quantum Monte Carlo as well as MPS techniques, see App.\ref{SubSM:Ground_state_calculations}), 
we find that both the FM and AFM experimental correlations are compatible 
with thermal ones at a temperature of $T/J\approx 0.35$.

\section{Friedel oscillations on an open ring}\label{Sec:Friedel}

\begin{figure}
	\centering
	\includegraphics{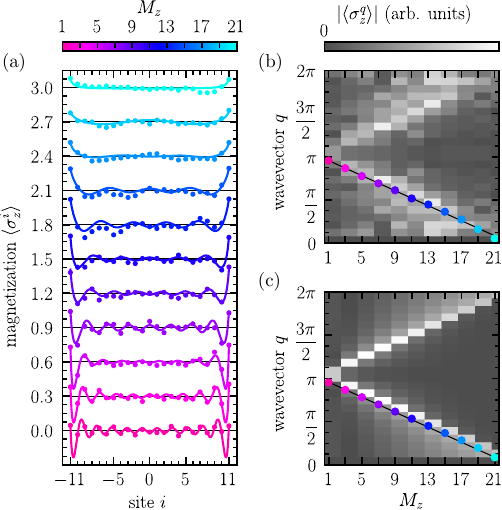}
	\caption{{\bf Friedel oscillations at the edges of a spin chain.}
		(a)~$z$-magnetization per site at the end of the adiabatic state 
		preparation of the AFM ground state, for various $M_z$ sectors.
		The $y$-axis corresponds to the magnetization of the open chain, 
		minus a background that was measured on the closed chain (see App.~\ref{SM:Friedel_oscillations}). 
		The curves are offset for clarity; the horizontal grey line displays the zero for each curve.
		Solid lines are fits to the data points using Eq.~(\ref{eq:Friedel}).
		(b)~Background: Fourier transform of the $z$-magnetization for each $M_z$ sector, 
		showing a linear shift of the spatial frequency. Colored circles: fitted 
		values of the oscillation frequency $2 k_{\rm F}$ using Eq.~(\ref{eq:Friedel}). 
		Solid black line: theory prediction given by Eq.~(\ref{eq:Friedel_k}), with no free parameter.
		(c)~Same as (b) based on a DMRG simulation of the open chain. The amplitude of the colormap in~(c) is larger by a factor~$2$ compared with the one in~(b).
	}
	\label{fig:fig3}
\end{figure}

The single-atom control of our experiment allows us to test another prediction of TLL theory, namely the 
existence of Friedel oscillations at the edges of an open chain \cite{Giamarchi_2004}. 
To do so, we modify the geometry of the chain by removing one atom, 
which amounts to considering a chain of $N=23$ spins with open boundary conditions (OBC) 
-- modulo the weak dipolar coupling across the hole.  
The AFM ground state is then expected to exhibit spatial oscillations of the local $z$-magnetization 
around the hole, akin to Friedel oscillations in fermionic systems~\cite{eggertMagneticImpuritiesHalfintegerspin1992,fathLuttingerLiquidBehavior2003}.
Examples of such behaviors have been observed 
with scanning tunneling microscopy at the edges of carbon nanotubes~\cite{Lee_2004} 
and near defects in WS$_2$ heterostructures~\cite{Li_2024}.
For an open chain with an odd number of atoms $N$ and a total magnetization $M_z$, 
the $z$-magnetization at a site $j$ away from the impurity takes the form (see
App.\ref{SM:Analytic_derivation_of_the_Friedel_oscillation}):
\begin{equation}\label{eq:Friedel}
	\langle \sigma_z^j \rangle \approx A \cos \left( 2 k_{\rm F} j \right) 
	\left[\frac{N}{\pi} \cos \left( \frac{\pi j}{N}\right)\right]^{-K(M_z)}~.
\end{equation}
Here $A$ is an overall amplitude, $K(M_z)$ is the magnetization dependent AFM Luttinger parameter, 
and $2 k_{\rm F}$ is the Friedel wavevector.
It obeys the kinematic relation
\begin{equation}\label{eq:Friedel_k}
	k_{\rm F} = \frac{\pi}{2} \left( 1 - \frac{M_z}{N} \right)~,
\end{equation}
which in the dual fermionic picture~\cite{Giamarchi_2004} is simply the distance between Fermi points as a function of filling.
Equivalently, it encodes a consistency condition between the microscopic symmetries (e.g. lattice-translation) 
of $H_{\rm{XY}}$ and the long-wavelength symmetries of the 
TLL~\cite{elseNonFermiLiquidsErsatz2021, chengLiebSchultzMattisLuttingerHooft2023}.

To check this prediction, we use the quasi-adiabatic preparation described above 
in order to drive the system close to the AFM ground state of the XY Hamiltonian 
on the open ring at fixed total magnetization $M_z$. 
To control the magnetization, we initialize $(N+M_z)/2$ spins in $\ket{\uparrow}$ (non-addressed atoms) 
and $(N-M_z)/2$ spins in $\ket{\downarrow}$ (addressed atoms), 
in a way that distributes as uniformly as possible the net magnetization
across the circle (see addressing patterns in App.~\ref{SM:Friedel_oscillations}). 
Then, we ramp down the light shifts $\delta$ acting on sublattice $B$ 
and measure the final state in the $z$ basis.

The resulting $z$-magnetization $\langle \sigma_z^i \rangle$ for each spin $i$ 
is plotted in Fig.~\ref{fig:fig3}(a), after subtracting a PBC background 
(see App.~\ref{SM:Background_subtraction_for_Friedel_oscillations}). 
We observe symmetric oscillations around the hole, 
consistent with the picture of Friedel oscillations. 
To probe their spatial frequency,  we Fourier-transform the $z$-magnetization profile: 
$\langle \sigma_z^q \rangle = \sum_j e^{i \, q j} \langle \sigma_z^j \rangle$ 
for $q \in \left\{ 2\pi n/N \right\}_{0\leqslant n < N}$. 
In Fig.~\ref{fig:fig3}(b), we plot $\left| \langle \sigma_z^q \rangle \right|$ 
as a function of the wavevector $q$ and the magnetization sector $M_z$. 
For each value of $M_z$, $\left| \langle \sigma_z^q \rangle \right|$ peaks at a 
given value of $q$; this value is consistent with the wavevector of the Friedel oscillation,
$2 k_{\rm F}$, linearly shifting from $\pi$ at small $M_z$ to $0$ at large $M_z$ 
[Eq.~(\ref{eq:Friedel_k})]. 
Extracting  $K(M_z)$ from the data turned out to be unreliable: while the dominant Friedel frequency is solely controlled by 
the magnetization per spin, the amplitude decay is additionally sensitive to e.g. residual finite energy 
due to lack of adiabaticity, and other experimental imperfections (see App.~\ref{SM:Background_subtraction_for_Friedel_oscillations}).

\section{Velocity of low-energy excitations}\label{Sec:Quench}

\begin{figure}
	\centering
	\includegraphics[width=\linewidth]{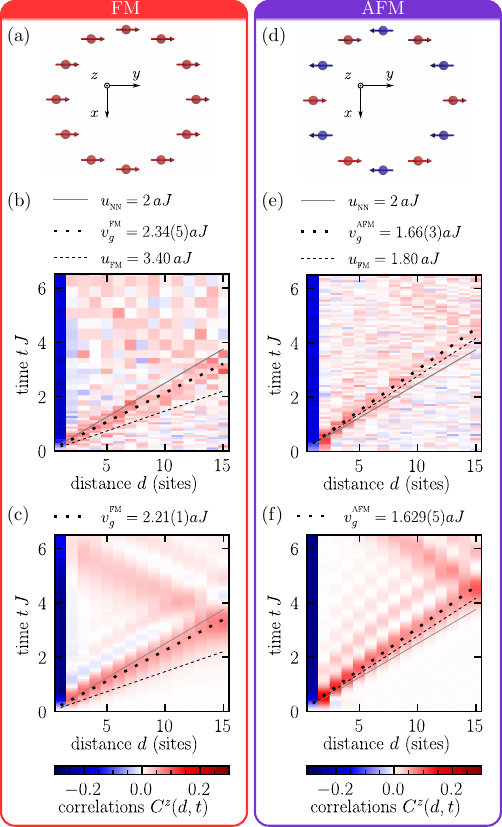}
	\caption{{\bf Measurement of the sound velocity $u$}, 
	by monitoring the propagation of correlations in quench experiments. 
	Data were taken with a chain of $N=30$ atoms using a 
	different mapping for the spins (see App.~\ref{SM:mapping_atoms_to_spins}).
	(a)~Sketch of the initial state used to measure the FM dispersion relation: 
	the spins are initialized in a coherent spin state along $y$ and evolve 
	freely under XY interactions.
	(b)~Evolution in space and time of the measured $z$-correlations $C^z (d,t)$. 
	The black dotted line shows $d=2 v_g^{_{\rm FM}} t$ where $v_g^{_{\rm FM}}$ 
	is extracted from a fit of the positive linear wavefront for distances $d>1$~site. 
	For comparison, the grey solid line shows $d = 2 u_{_{\rm NN}} t$ with $u_{_{\rm NN}}=2 a J$ 
	the expected sound velocity assuming only nearest neighbor interactions. 
	The dashed line shows the predicted sound velocity $u_{_{\rm FM}}$ of the TLL theory.
	(c)~Evolution in space and time of the simulated $z$-correlations, 
	with no free parameter (see App.~\ref{SM:Numerical_methods}).
	(d,e,f)~Same as (a,b,c) in the AFM case.}
	\label{fig:fig4}
\end{figure}

We conclude our study by probing the low-energy excitations of $\pm H_{\rm XY}$, in order to  
measure the second parameter of the TLL Hamiltonian, namely the sound velocity~$u$. 
Following a method we have used in 2D \cite{Chen_2023_2}, 
we initialize the system in a low-energy product state which is not an eigenstate of $H_{\rm XY}$, 
and monitor its free evolution under $H_{\rm XY}$. This is similar to what was done in 
Ref.~\cite{Yang_2017} for a 1D Bose gas, and alternative to e.g. Bragg spectroscopy of 1D gases \cite{Yang_2018}. 
For the initial state we choose the coherent spin state 
$\ket{\rm CSS} = \ket{\rightarrow_y \rightarrow_y \cdots \rightarrow_y}$ where all spins are 
pointing along $y$ [Fig.~\ref{fig:fig4}(a)].
This state is a mean-field approximation of the FM ground state of $H_{\rm XY}$. 
We then let the system evolve for a time $t$, and measure the $z$-correlations. 
The resulting evolution of $C^z(d,t)$ is shown is Fig.~\ref{fig:fig4}(b) 
where $d$ is the perimeter distance along the ring. 
We observe two distinct behaviors depending on the distance: 
first, nearest-neighbor correlations ($d=1$) build up in less than $1/J$ and reach a quasi-stationary negative value; 
second, for $d>1$ a positive correlation wavefront spreads ballistically from $d=2$ up to the largest distances. 

The ballistic propagation occurs at a velocity $2v_g$, revealing the characteristic group 
velocity $v_g$  of the excitations of the system \cite{Cheneau_2012}. 
A two-dimensional fit of the data gives $v_g = 2.34(5) a J$, which is larger than the sound 
velocity of a system with NN interactions $u_{_{\rm NN}} = 2 a J$. 
This reveals that dipolar FM interactions accelerate the dynamics compared to NN ones. 
Also, contrary to the 2D case \cite{Chen_2023_2} or to 1D systems with longer-range interactions   
\cite{Jurcevic_2014, Richerme_2014}, they maintain a linear light-cone dynamics, as expected for a TLL. 
Yet the sound velocity expected theoretically for the dipolar FM chain $u_{\rm FM} \approx 3.7 a J$
is significantly larger than the measured one. 
This discrepancy can be explained by inspecting 
the whole spectrum of excitations, which reveals that the linear sound mode involves only 
very few wavevectors in our small ring (see App.\ref{SM:Luttinger}).
As our quench protocol does not populate selectively those few wavevectors, 
the correlation dynamics is instead dominated by modes at intermediate wavevectors, 
which possess a smaller effective group velocity. 

To study the correlation dynamics in the AFM chain, we perform the same protocol up to a sublattice
rotation~\cite{Chen_2023_2}: as shown in Fig.~\ref{fig:fig4}(e), the initial state 
is now a staggered coherent spin state 
$\ket{\rm CSS}_{\rm stag} = \ket{\rightarrow_y \leftarrow_y \cdots \rightarrow_y  \leftarrow_y}$, 
which is the mean-field ground state of the AFM Hamiltonian $-H_{\rm FM}$. 
The qualitative behavior of $C^z(d,t)$ is the same, but the propagation of correlations 
is now slower than the one with NN interactions: $v_g= 1.66(3)  aJ$, 
a signature of frustration of dipolar AFM interactions.  
This time the observed light-cone velocity is much closer to the sound velocity 
predicted theoretically,  $u_{\rm FM} \approx 1.8 a J$, reflecting the fact that in the 
AFM case the linear sound mode dominates the spectrum at small wavevectors  (see App.\ref{SM:Luttinger}).

\section{Conclusion}\label{Sec:Conclusion}

In this work, we have shown that a 1D chain of Rydberg-encoded spins interacting 
under the dipolar XY Hamiltonian realizes the physics of one-dimensional gapless spin liquids, 
belonging to the family of Tomonaga-Luttinger liquids.
The measured FM and AFM correlation profiles reveal the expected power-law decays, 
and the discrepancies with the theoretical ground-state correlations can be 
explained by experimental imperfections. 
We have observed Friedel oscillations around an impurity and have verified the 
expected dependence of the oscillation wavevector with the magnetization. 
Finally, the quench dynamics at low energy shows a linear light-cone propagation of correlations, 
which allows us to extract the sound velocity of a 
Tomonaga-Luttinger liquid when the sound mode dominates the low-energy spectrum. 

Our analysis is based on the reconstruction of correlations in real space and real time for a small 
sample of synthetic matter. Our work probes the robustness of LL physics 
to finite-size effects and to various imperfections
such as the presence of holes. Controlling all these effects in future experiments 
will then allow for  more stringent tests of one-dimensional quantum physics.
It thus offers a complementary view on TLLs compared to 
experiments performed on bulk materials, such as spin-chain compounds. 
The methods that we use are quite general, and could be extended 
to study other many-body effects such as the transition from TLLs to 
phases dominated by disorder \cite{Giamarchi_1988,Fisher_1994,Altman_2004,D'Errico_2021}, 
and the correlations and Friedel oscillations in two-dimensional gapless spin liquids~\cite{Bintz_2024}.

\begin{acknowledgments}
	We acknowledge the insightful discussions with Isabelle Bouchoule, Thierry Giamarchi, 
	Edmond Orignac and Guido Pupillo.
	This work is supported by
	the Agence Nationale de la Recherche (ANR-22-PETQ-0004 France 2030, project QuBitAF), 
	the European Research Council (Advanced grant No. 101018511-ATARAXIA), and 
	the Horizon Europe programme HORIZON-CL4- 2022-QUANTUM-02-SGA (project 101113690 (PASQuanS2.1).
	Numerical calculations were performed on the CBPmsn cluster at the ENS of Lyon and on the FASRC Cannon cluster supported by the FAS Division of Science Research Computing Group at Harvard University.
	M.B. acknowledges support from the NSF via the Harvard-MIT Center for Ultracold Atoms.
	N.Y.Y acknowledges support from the U.S. Department of Energy via the QuantISED 2.0 program and from a Simons Investigator award. 
	D.B. acknowledges support from MCIN/AEI/10.13039/501100011033 (PID2020-119667GA-I00, 
	CNS2022-13578, EUR2022-134067 and European Union NextGenerationEU PRTR-C17.I1).
	
\end{acknowledgments}


\setcounter{figure}{0}
\renewcommand\thefigure{A\arabic{figure}} 
\appendix


\section{Mapping of atomic states onto a spin model} \label{SM:mapping_atoms_to_spins}

All experiments were performed with arrays of $^{87}$Rb atoms trapped in 
optical tweezers, using the setup described in previous works~\cite{Nogrette_2014, Barredo_2016,Chen_2023}.
The mapping from two Rydberg states onto a spin 1/2 is: 
$\ket{\uparrow} = \ket{n S_{1/2},m_J=1/2}$ and  $\ket{\downarrow} = \ket{n P_{1/2},m_J=-1/2}$ 
with $n$ the principal quantum number. A $45$~G magnetic field 
perpendicular to the array ensures isotropic interactions.
Spin rotations are performed using  microwave pulses with Gaussian temporal envelope and 
the magnetic field  isolates the effective spin states from other Zeeman levels.

The atoms in those states interact under the dipole-dipole Hamiltonian, 
which in spin language translates into the following Hamiltonian~\cite{Browaeys_2020}:
\begin{equation}\label{eq:Hamiltonian_SM}
	H_{\rm tot} = H_{\rm XY} + H_{\rm vdW}.
\end{equation}
Here, $H_{\rm XY}$ is the first-order contribution of effective Hamiltonian theory 
and is given by Eq.~(\ref{eq:Hamiltonian}) of the main text; 
the second-order van der Waals (vdW) contribution $H_{\rm vdW}$ reads
\begin{equation}\label{eq:Hamiltonian_vdW_SM}
	H_{\rm vdW} =
	\sum_{i < j}  \frac{1}{r_{ij}^6}  \sum_{(st) \in \{\uparrow, \downarrow\}^2} U_{s,t} \; n_i^s n_j^t
\end{equation}
with $n_i^{\uparrow}=(1 + \sigma_i^z)/2$ and $n_i^{\downarrow}=(1 - \sigma_i^z)/2$. 
The vdW Hamiltonian leads to small corrections compared 
to the pure XY Hamiltonian. It is taken into account in all our numerical simulations. 
The values of the interaction energies are estimated from~\cite{Weber_2017} 
and are summarized in Table~\ref{tab:interaction_energies}.
The vdW term can be written in terms of spin operators as
$H_{\rm vdW} = - J_{\rm vdW}/2 \sum_{i<j} (1/r_{ij}^6) \sigma_i^z \sigma_j^z$, 
up to uniform field terms which are irrelevant when the magnetization is conserved. 

The choice of $n$ results from a trade-off between having the longest possible Rydberg lifetimes 
(scaling as $n^3$) and a not-too-large XY interaction energy $J$ (scaling as $n^4$) 
to make initial-state preparation and detection easier (see details in Sec.~\ref{SM:Experimental_errors}). 
For the adiabatic preparation schemes, we choose $n=70$ in a $N=24$~atoms 
chain with nearest-neighbor distance $16.2$~$\mu$m, 
whereas for the quench experiments we choose $n=60$ to have more atoms ($N=30$) 
in the same area, with a similar interaction energy (n.n. distance $13\,\mu$m).

\begin{table*}[]
	\centering
	\begin{tabular}{c|c|c}
		Sequence                           & Adiabatic preparation & Quench experiment \\ \hline \hline
		Principal quantum number & $n=70$ & $n=60$ \\ \hline
		Nearest-neighbor distance & $16.2$~$\mu$m & $13$~$\mu$m \\ \hline
		XY interaction energy [Eq.~(\ref{eq:Hamiltonian})] & $J=2\pi\times0.55$~MHz & $J=2\pi\times0.62$~MHz \\ \hline
		van der Waals interaction energies [Eq.~(\ref{eq:Hamiltonian_vdW_SM})] & \begin{tabular}[c]{@{}l@{}}
			$U_{\uparrow,\uparrow}=2\pi\times0.051$~MHz\\ $U_{\downarrow,\downarrow}=-2\pi\times0.007$~MHz\\ 
			$U_{\uparrow,\downarrow}=U_{\downarrow,\uparrow}=2\pi\times0.058$~MHz
		\end{tabular} &
		\begin{tabular}[c]{@{}l@{}}
			$U_{\uparrow,\uparrow}=2\pi\times0.030$~MHz\\ 
			$U_{\downarrow,\downarrow}=-2\pi\times0.006$~MHz\\ $U_{\uparrow,\downarrow}
			=U_{\downarrow,\uparrow}=2\pi\times0.009$~MHz
		\end{tabular}\\
	\end{tabular}
	\caption{Values of the interaction energies for the various experimental sequences, for a $45$~G magnetic field 
	perpendicular to the array to guarantee isotropic interactions.}
	\label{tab:interaction_energies}
\end{table*}

\begin{figure*}
\centering
\includegraphics{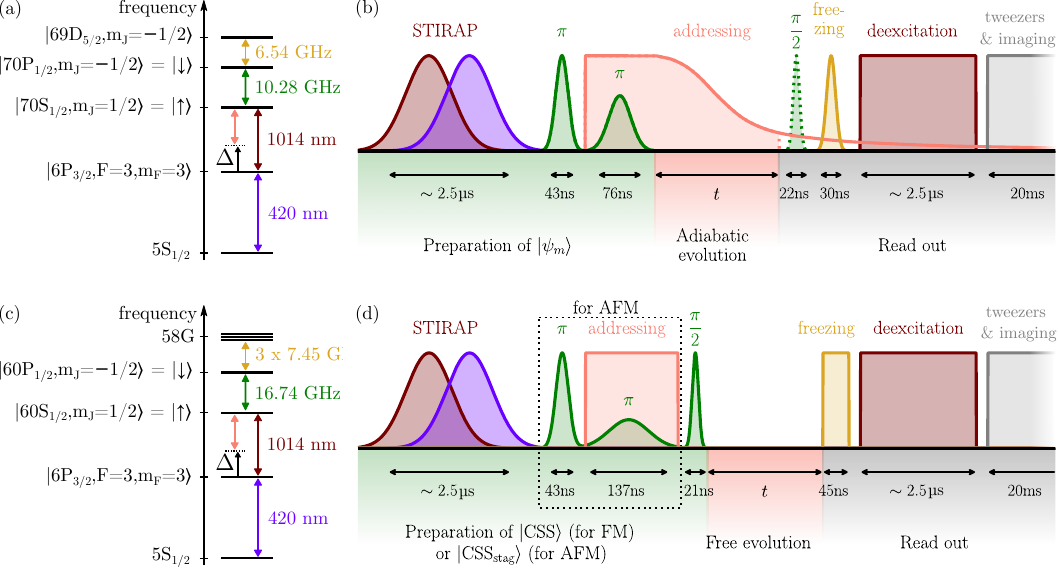}
\caption{\textbf{Experimental sequence for adiabatic preparation of XY 
ground states (Sec.~\ref{Sec:GS} and \ref{Sec:Friedel} of the main text) 
and for the quench experiment (Sec.~\ref{Sec:Quench}).}
(a)~Atomic energy levels involved in the adiabatic sequence and associated transitions: 
$420$~nm and $1014$~nm are the wavelengths used for Rydberg excitation, 
and $10.28$~GHz and $6.54$~GHz are microwave fields for driving Rydberg-Rydberg transitions.
(b)~Sketch of the experimental sequence for adiabatic preparation of the XY FM and AFM 
ground state of $\pm H_{\rm XY}$. For clarity the timings of the pulses are not to scale, 
and their colors refer to the transition energies in (a). 
The read-out depends on the basis of measurement: to measure spins in the $z$-basis, 
we let the addressing light shifts on (pink solid line), freeze the interaction dynamics (yellow pulse) 
and read the atomic state of each atom (red and grey pulses); 
to measure the spins in the $xy$-plane, we switch off the addressing light shifts 
(pink dotted line) and perform a global microwave $\pi/2$ pulse (green dotted line), 
before freezing and reading of the atomic state.
(c)~Atomic energy levels involved in the quench sequence.
(d)~Sketch of the experimental sequence for the quench experiment from the 
XY mean-field ground state of $\pm H_{\rm XY}$. 
The dotted black rectangle shows the part of the sequence which is specific to the AFM case.
}
\label{fig:SM_sequence}
\end{figure*}

\section{Experimental sequence} \label{SM:exp_sequence}

The beginning of the experimental sequence, as well as the read-out, 
are common to all experiments. 
Starting from a cloud of $^{87}$Rb atoms in a magneto-optical trap, 
single atoms are loaded into optical tweezers 
and rearranged into a defect-free array with the desired geometry~\cite{Schymik_2020}. 
Atoms are Raman sideband cooled down to $\sim10$~$\mu$K, 
and optically pumped into the state $\ket{5S_{1/2},F=2,m_F=2}$. 
After that, we adiabatically ramp down the tweezers power to reduce the velocity 
dispersion of the atoms; next, we switch off the tweezers, and excite all atoms to 
the Rydberg state $\ket{\uparrow}$ using a stimulated Raman adiabatic 
passage (STIRAP) via the intermediate state $\ket{6P_{3/2}, F=3,m_F=3}$.
The following of the sequence depends on the type of experiment 
that we perform: either an adiabatic preparation of XY ground states 
(Sec.~\ref{Sec:GS} and \ref{Sec:Friedel} of the main text), 
or a quench experiment from the mean-field ground state (Sec.~\ref{Sec:Quench}).

At the end of the sequence, we perform a projective measurement of 
each atom's state, and repeat it more than $1000$ times with the same sequence 
to acquire statistics. Each projective measurement consists of four steps.
\begin{enumerate}
	\item A global microwave pulse defines the measurement basis. 
	In the absence of this pulse, the measurement basis is $z$. 
	To measure spins in the $x y$ plane, we apply a global $\pi/2$ pulse, 
	whose phase determines the basis ($x$ or $y$ or any combination of those bases).
	
	\item To prevent the spins from evolving during the following of the read-out process, 
	the spin dynamics is stopped by a ``freezing" pulse 
	which removes the atoms in $\ket{\downarrow}$ faster than the typical evolution time 
	$2\pi/J \sim 1$~$\mu$s. Depending on the atomic states used in the mapping, 
	we send the atoms either to the state $\ket{69 D_{5/2},m_J=-1/2}$ 
	with a single-photon Gaussian pulse [Fig.~\ref{fig:SM_sequence}(a,b)], 
	or to the hydrogenic manifold $58G$ via a three-photon square pulse [Fig.~\ref{fig:SM_sequence}(c,d)].
	
	\item Atoms in $\ket{\uparrow}$ are deexcited to the ground state manifold $5S_{1/2}$ 
	by shining a pulse of $1014$~nm light on resonance with the short-lived state $\ket{6P_{3/2},F=3,m_F=3}$.
	
	\item Tweezers are switched back on to recapture the atoms in $5S_{1/2}$ and eject those 
	remaining in the Rydberg states by the ponderomotive force. Finally, we perform a global 
	fluorescence imaging of atoms in $5S_{1/2}$, and map the imaged atoms on $\ket{\uparrow}$, 
	whereas the lost atoms are considered as $\ket{\downarrow}$.
\end{enumerate}

\section{Experimental methods for adiabatic preparations} \label{SM:method_adiabatic_preparation}

To prepare the ground state of $\pm H_{\rm XY}$, we use the same adiabatic protocol as in~\cite{Chen_2023}. 
The sequence is shown in Fig.~\ref{fig:SM_sequence}(b). 
A first global microwave $\pi$ pulse transfers all atoms from $\ket{\uparrow}$ to  $\ket{\downarrow}$. 
Then, we switch on the addressing light shifts on sublattice $B$, 
and perform another $\pi$ pulse on resonance with the atoms of sublattice $A$ only, 
thus preparing the state $\ket{\psi_m}$ with $(N+m)/2$ spins in $\ket{\uparrow}$ 
(sublattice $A$) and $(N-m)/2$ spins in $\ket{\downarrow}$ (sublattice $B$). 
Next, we ramp down the addressing light shifts $\delta(t)$ by controlling the 
addressing optical power with an acousto-optic modulator (AOM). 
In the particular case where the energy gap between the ground state and the first excited 
state depends linearly on $\delta$, one can derive the following analytical expression 
for the adiabatic ramp (up to the response time of the AOM)~\cite{Richerme_2013,Fang_2024}:
\begin{equation} \label{eq:LILA_generic}
	\delta(t) = \frac{E_0 \delta_c t + E_c \delta_0 (T-t)}{E_0 t + E_c (T-t)},
\end{equation}
with $T$ the duration of the ramp, $\delta_0$ the initial light shift, $\delta_c$ the light shift at the critical point, 
$E_0$ the energy gap for $\delta = \delta_0$ and $E_c$ the gap for $\delta = \delta_c$. 
For the 1D XY model, the critical point occurs at $\delta_c=0$, so that Eq.~(\ref{eq:LILA_generic}) can be simplified into
\begin{equation} \label{eq:LILA_specific}
	\delta(t) = \delta_0 \frac{T-t}{T-(1-\alpha)t}
\end{equation}
where $\alpha = E_0 / E_c$ 
($E_c$ is a finite-size energy gap, expected to vanish in the thermodynamic limit). 
The values of $T$ and $\alpha$ were optimized empirically to maximize the 
experimentally-measured $x$-correlations, and the data from the main text was taken with: 
$(T,\alpha)=(1.5~$$\mu$s$,20)$ in the FM case, and $(T,\alpha)=(2.5~$$\mu$s$,100)$ 
in the AFM case. The value of $\alpha$ is larger in the AFM case, in agreement with 
the fact that the critical gap $E_c$ is smaller in the AFM case, 
due to the weak frustration induced by next-nearest neighbor couplings.

\section{Experimental sequence for the quench experiment}

For the quench experiment discussed in Sec.~\ref{Sec:Quench}, 
we use the same protocol as in~\cite{Chen_2023_2}. 
The sequence is summarized in Fig.~\ref{fig:SM_sequence}(d), 
and it differs for FM and AFM. In the FM case, a global microwave $\pi/2$ 
pulse after the STIRAP directly prepares the targeted state 
$\ket{\rm CSS} = \ket{\rightarrow_y \rightarrow_y \cdots \rightarrow_y}$. 
In the AFM case, we add a set of microwave and addressing pulses 
[shown inside the black dotted frame of Fig.~\ref{fig:SM_sequence}(d)] 
to first prepare the state $\ket{\psi_0} = \ket{\uparrow \downarrow \cdots \uparrow  \downarrow}$; 
then we rotate it with a global $\pi/2$ pulse, thus preparing the state 
$\ket{\rm CSS}_{\rm stag} = \ket{\rightarrow_y \leftarrow_y \cdots \rightarrow_y  \leftarrow_y}$. 
The initial state preparation is followed by a free evolution under $H_{\rm tot}$.

\section{Experimental imperfections}\label{SM:Experimental_errors}

Several imperfections limit the fidelities of the previously described adiabatic protocol.

\subsection{Finite fidelity of the initial state preparation}

The initial state preparation suffers from two limitations.
First, the Rydberg excitation has a finite efficiency $1-\eta_{\rm STIRAP}$, 
leading to a small portion of atoms $\eta_{\rm STIRAP} \approx 2\%$ 
which are left in the ground state and do not interact with the other atoms.
Second, the spin rotations are performed in the presence of interactions and 
with finite light shifts, which reduce their efficiency by $\sim 1-2 \%$ and 
thus increase the energy of the initial state compared 
with the ideal ground state of $H_{\rm Z}$.

\subsection{Non-adiabaticity and decoherence}

Several decoherence phenomena affect the time evolution of the system, especially the adiabaticity.

A major limitation is the finite Rydberg lifetimes, which induce leakages 
from the ideal isolated two-level system $\left(\ket{\uparrow},\ket{\downarrow}\right)$ 
to other atomic states. There are two contributions to the Rydberg lifetimes~\cite{Gallagher_1994}: 
spontaneous emission to the low-lying energy states, at a rate $1/\tau_0$; depopulation 
by absorption and stimulated emission of black-body radiation in our room-temperature setup, 
at a rate $1/\tau_{\rm BRR}$. Those contributions add up to an effective 
decay rate $1/\tau_{\rm eff} = 1/\tau_0 + 1/\tau_{\rm BRR}$ that depends on the 
considered Rydberg level. For $n=70$, we 
estimate~\cite{Beterov_2009}: 
$\tau_{\rm eff}^{\ket{\uparrow}}\approx144$~$\mu$s and $\tau_{\rm eff}^{\ket{\downarrow}}\approx172$~$\mu$s.

The lifetime of the state $\ket{\uparrow}$ can also be reduced in the presence of the addressing, 
which can induce transitions to the short-lived state $6P_{3/2}$ that then decays to the ground 
state manifold $5S_{1/2}$. We limited this depumping effect by initializing the addressed atoms 
in the state~$\ket{\downarrow}$, such that the depumping can only happen during the adiabatic ramp, 
when the addressing light shifts are still on and the addressed atoms are partially in $\ket{\uparrow}$. 
We checked experimentally that the percentage of depumped atoms 
due to the addressing is smaller than 0.5~\% at the end of the adiabatic ramp. 

Another decoherence effect comes from the fluctuations of the atomic positions. 
During the Rydberg sequence, atoms are in free flight: 
the standard deviation of their position along $\mu$ 
increases as $\sigma_\mu^{\rm tot}(t) = \sqrt{\sigma_\mu^2 +\left( \sigma_{v_\mu} t \right)^2}$, 
with $\sigma_\mu$ the position uncertainty of the atoms in their the tweezer, 
and $\sigma_{v_\mu}$ their velocity uncertainty. 
We estimate that in the radial directions, $\sigma_{x,y} \sim 100$~nm and 
$\sigma_{v_{x,y}} \sim 25$~nm/$\mu$s, whereas in the axial direction 
$\sigma_z \sim 800$~nm and $\sigma_{v_z} \sim 40$~nm/$\mu$s. 
This leads to fluctuations of the couplings between spins,
which are averaged over many slightly disordered geometries.

\begin{figure}
	\centering
	\includegraphics[width=\linewidth]{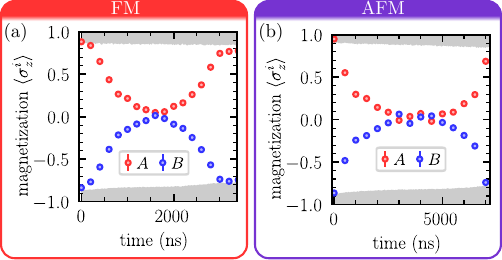}
	\caption{\textbf{Adiabatic preparation of XY ground states and back.}
	Time evolution of the magnetization per sublattice in the FM case (a) and in the AFM case (b), 
	in an experiment where light shifts are ramped down and then ramped up in a symmetric way. 
	The ramp up starts at $t=1.5$~$\mu$s in the FM case and at $t=2.5$~$\mu$s in the AFM case. 
	The grey regions indicate the expected loss of contrast from single-atom imperfections, 
	due to finite Rydberg lifetimes and depumping of addressed atoms (see text), 
	starting from the experimentally-measured contrast.
	}
	\label{fig:SM_ramp_down_ramp_up}
\end{figure}

The level of adiabaticity can be quantified by performing a back-and-forth adiabatic preparation: 
starting from the state $\ket{\psi_0}$, we first ramp down $\delta$ to $0$, 
and then ramp it up in a symmetric way to get back to the initial state $\ket{\psi_0}$. 
Fig.~\ref{fig:SM_ramp_down_ramp_up} shows the resulting $z$-magnetization 
in both FM and AFM cases. Whereas we would ideally expect the $z$-magnetization 
to come back to its initial value, we observe a reduction of the contrast of about 
$12\%$ in the FM case (from $86\%$ at $t=0$ to $76\%$ at $t=3.4$~$\mu$s), 
and about $21\%$ in the AFM case (from $90\%$ at $t=0$ to $71\%$ at $t=7.3$~$\mu$s).

The observed loss of contrast may come from adiabatic imperfections, 
but also from single-particle errors: finite Rydberg lifetimes and depumping of addressed atoms. 
To estimate those contributions, we perform Monte-Carlo simulations for the recapture probabilities 
of addressed and non-addressed atoms: first, the atomic state is randomly sampled according to 
the calibrated preparation errors; then its time evolution is computed 
accounting only for {\it single-particle} errors. The results are shown as the grey regions in 
Fig.~\ref{fig:SM_ramp_down_ramp_up}. We find that single-atom errors explain 
half of the contrast reduction, the remaining being the adiabatic imperfections.

\subsection{Read-out errors}

Spin states can be misread during the measurement phase. 
A spin $\ket{\uparrow}$ has a probability $\varepsilon_{\uparrow} = 2.5 \pm 1 \%$ 
to be measured as $\ket{\downarrow}$, owing to mechanical losses and finite deexcitation efficiency. 
Conversely, a spin $\ket{\downarrow}$ has a probability $\varepsilon_{\downarrow} = 3 \pm 1 \%$ 
to be measured as $\ket{\uparrow}$ due to spontaneous emission to $5S_{1/2}$ before the atom 
is kicked out from the tweezers. The effect of detection errors on the measured observables can be computed. 
The average magnetization $\langle \sigma_{\mu}^i \rangle$ and correlations $C_{i,j}^{\mu}$ 
are related to the same quantities $\langle \tilde{\sigma_{\mu}^i} \rangle$ and $\tilde{C}_{i,j}^{\mu}$ 
without detection errors by the relations (valid to first order in $\varepsilon_{\uparrow, \downarrow}$):
\begin{align}
	\langle \sigma_{\mu}^i \rangle &= \left(1-\varepsilon_{\downarrow}-\varepsilon_{\uparrow}\right) 
	\langle \tilde{\sigma_{\mu}^i} \rangle + \varepsilon_{\downarrow} - \varepsilon_{\uparrow}\\
	C_{i,j}^{\mu} &= \left(1-2\varepsilon_{\downarrow}-2\varepsilon_{\uparrow}\right) \tilde{C}_{i,j}^{\mu}.
	\label{eq:effect_of_detection_errors_on_correlations}
\end{align}
These expressions can be inverted to correct magnetization and correlations from detection errors. 

\section{\texorpdfstring{$U(1)$ symmetry of the prepared states}{U(1) symmetry of the prepared states}}\label{SM:symmetry}

\begin{figure}[h!]
	\centering
	\includegraphics[width=\linewidth]{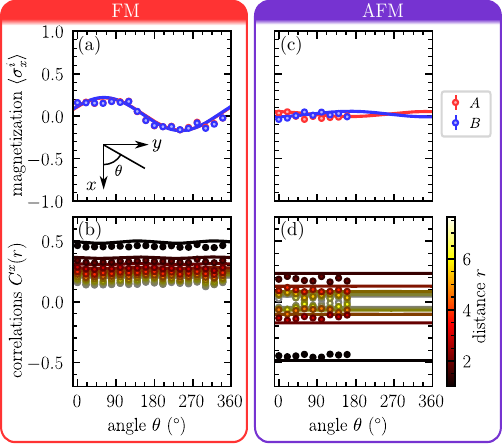}
	\caption{\textbf{Angular dependence of the magnetization and correlations in the $xy$ plane.}
		(a)~Magnetization per sublattice, as a function of the angle $\theta$ in the $xy$ plane (inset), 
		in the FM case, measured at $t=1.6$~$\mu$s. The variation of the angle is obtained 
		by scanning the phase of the measurement pulse.
		(b)~Correlations as a function of $\theta$, in the FM case, at $t=1.6$~$\mu$s.
		(c,d)~Same as (a,b) in the AFM case, at $t=3.5$~$\mu$s.
	}
	\label{fig:SM_scan_phase}
\end{figure}

In this section, we check if the states prepared using the adiabatic protocol 
of Sec.~\ref{Sec:GS} satisfy the expected $U(1)$ symmetry of $H_{\rm XY}$. 
For this purpose, we perform a measurement at the end of the ramp along a given direction 
$\cos(\theta) x + \sin(\theta) y$, and we scan the angle $\theta$. 
The resulting magnetization $\langle \sigma_{\theta}^i \rangle$ shows a residual 
oscillation [Fig.~\ref{fig:SM_scan_phase}(a,e)], which we attribute to the repercussion of a 
small $U(1)$ symmetry breaking in the initial state. 
The amplitude of the oscillation is larger in the FM case, in agreement with the numerical simulations.
However, the connected correlations $C^\theta (r)$ are isotropic up to the statistical noise, 
and simulations confirm the weak dependence of $C^\theta (r)$ with $\theta$ [Fig.~\ref{fig:SM_scan_phase}(b,f)]: 
The connected character of the correlations tends to compensate the symmetry-breaking observed in the magnetization.

\section{Fitting procedure for the correlation profiles}\label{SM:fitting_procedure}

Here, we explain our protocol to extract the power-law exponent of the correlations 
from their spatial profile shown in Fig.~\ref{fig:fig2} of the main text. 
This procedure may suffer from three biases:
\begin{enumerate}
	\item At large distances, any finite energy density turns the power-law decay into an exponential decay.
	
	\item At short distances, the theoretical expression from the LL theory 
	[Eqs.~(\ref{eq:Cx_FM},\ref{eq:Cz_FM}) of the main text] may not describe 
	accurately the system, since this field theory is only valid at asymptotic distances $r \gg 1$.
	
	\item Independent detection errors rescale the correlation profile by a global multiplicative factor 
	$1- 2\varepsilon_{\downarrow} - 2\varepsilon_{\uparrow} = 0.89$, whatever the distance between 
	the spins [Eq.~(\ref{eq:effect_of_detection_errors_on_correlations})]. 
	This does not modify the power-law exponent of the correlations, 
	but it affects their amplitude which depend on $K$ in the case of the $z$-correlations [Eq.~\ref{eq:Cz_FM}].
\end{enumerate}

To limit the large-distance bias, we choose to incorporate a finite correlation length~$\xi$ 
into our fitting functional form, as already explained in the main text: 
$\tilde{C}^x(r) = C^x(r) e^{-r/\xi}$, in both FM and AFM cases. 
This is done only along~$x$, since the measured $z$-correlations appear less 
sensitive to finite-temperature effects.

\begin{figure}[h!]
	\centering
	\includegraphics[width=\linewidth]{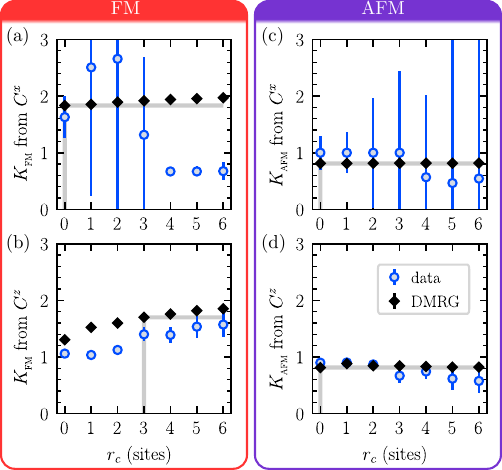}
	\caption{\textbf{Effect of a lower cutoff on the fitted Luttinger parameter~$K$}.
	Each panel corresponds to one of the four spatial correlation profiles shown in 
	Fig.~\ref{fig:fig2} of the main text: 
	$C^x(r)$ in the FM case~(a) and in the AFM case~(c); 
	$C^z(r)$ in the FM case~(b) and in the AFM case~(d). 
	For each correlation profile, we fit the correlations for distances $r \geq r_c$, 
	for both the ideal ground state (simulated with DMRG, black diamonds) and the data 
	(blue points), and we repeat the fit for various cutoff distances~$r_c$. 
	The chosen value of $r_c$ is indicated by the grey vertical line.
	}
	\label{fig:SM_scan_cutoff_distance}
\end{figure}

To avoid the second bias, we define a cutoff distance $r_c$ and fit correlations $C(r)$ 
on distances $r \leq r_c$. We set the value of $r_c$ by repeating the fits for 
different values of $r_c$ on the ideal ground state (simulated by DMRG), 
and choose the smallest cutoff that gives a satisfying convergence. 
This analysis is shown in Fig.~\ref{fig:SM_scan_cutoff_distance}. 
The only case which is sensitive to the cutoff is the FM correlations along~$z$, 
for which we set $r_c=3$~sites; in the three other cases we set $r_c=0$.

Finally, to take into account the effect of independent detection errors, 
we included the rescaling factor due to the detection errors in the fitting functional 
form for the $z$-correlations: 
$\tilde{C}^z(r) = \left(1- 2\varepsilon_{\downarrow} - 2\varepsilon_{\uparrow}\right) C^z(r)$.

\section{Numerical methods}\label{SM:Numerical_methods}

\subsection{Simulation of the adiabatic ramp with imperfections}
\label{SubSM:Simulation_of_the_adiabatic_ramp_with_imperfections}

\begin{figure}
	\centering
	\includegraphics[width=\linewidth]{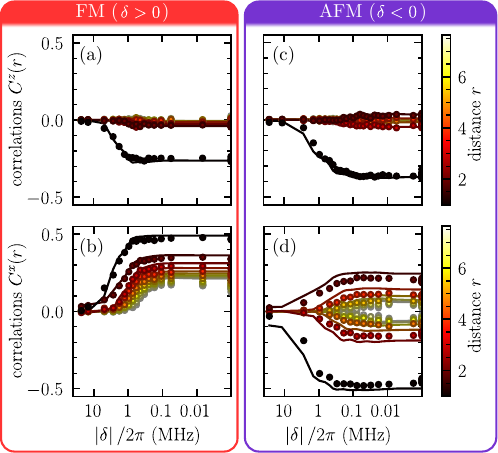}
	\caption{\textbf{Comparison of the measured correlations (data points) 
	with numerical simulations of the dynamics (solid lines).} 
	The first row (a,c) is the correlations along $z$ for all distances, and the second row is the correlations along $x$.
	}
	\label{fig:SM_data_VS_TDVP}
\end{figure}

The dynamical simulations of the ramp were performed using the time-dependent 
variational principle (TDVP) with matrix product states (MPS), 
implemented via the ITensor package~\cite{Fishman_2022} with truncation error $< 10^{-10}$ 
and bond dimension up to 200. The simulations contain no free parameter.

The state following the laser Rydberg excitation consists of all spins pointing up, 
where each spin is replaced by a hole with a probability of 0.02\% to describe the STIRAP errors. 
A hole represents a spin that does not interact with other spins and thus does not contribute to the observables. 
Moreover, during the time evolution, new holes appear due to the finite lifetime of the Rydberg states. 
They are incorporated into the simulation using the quantum trajectory method~\cite{Daley_2014} 
with four distinct decay channels: 
transitions from the $\ket{\uparrow}$ or $\ket{\downarrow}$ states to either the atomic ground state $\ket{\text{g}}$ 
or another Rydberg state $\ket{\text{r}'}$. 
Consequently, the local Hilbert space is expanded to $\text{dim}(\mathcal{H}_i)=4$.
We introduce $4N$ jump operators $c_{i,m}$, where $m=1,\dots,4$ defined as
\begin{equation}
	\begin{split}
		c_{i,1} & = \sqrt{\gamma_1} \ket{\text{g}}_i  \bra{\uparrow} \ , \
		c_{i,2} = \sqrt{\gamma_2}   \ket{\text{r}'}_i \bra{\uparrow} \\
		c_{i,3} & = \sqrt{\gamma_3} \ket{\text{g}}_i  \bra{\downarrow} \ , \
		c_{i,4} = \sqrt{\gamma_4}   \ket{\text{r}'}_i \bra{\downarrow},
	\end{split}
\end{equation}
where $\gamma_{i,m}$ represents the decay rate associated with the decay channel $m$ for the spin $i$.
We evaluate at each step of the evolution 
the probability associated to every decay channel 
$\delta p_{i,m} = \delta t \bra{\Psi(t)} \gamma_{i,m} c^\dag_{i,m} c_{i,m} \ket{\Psi(t)}$,
where $\ket{\Psi(t)}$ denotes the state at time $t$.
Calling $\delta p = \sum_{i,m} \delta p_{i,m}$ the probability that a quantum jump occurs, 
the state is stochastically evolved with probability $\delta p$ to
$\ket{\Psi(t+\delta t)} = \sqrt{\gamma_{i,m}} \ c_{i,m} \ket{\Psi(t)}$
where one particular channel is chosen with probability $\delta p_{i,m}/ \delta p$, otherwise to
$\ket{\Psi(t+\delta t)} = \exp[- i H_{\text{eff}} \delta t ] \ket{\Psi(t)}$,
with the effective Hamiltonian given by
$H_{\text{eff}} = H - \frac{i}{2} \sum_{i,m} \gamma_m \ c^\dag_{i,m} c_{i,m}$.
We assumed state- and site-independent decay rates $\gamma_m=0.0037$. 
This significantly simplifies the picture, as the non-Hermitian Hamiltonian is then proportional to the identity.

We also simulate the microwave pulses to 
prepare the N\'eel state and to measure observables in the $x,y$ plane. 
During those pulses, the spins still interact via $H_{\rm tot}$.
For a meaningful comparison of observables in the $xy$ plane between experiment and theory, 
we need to align the $x$ and $y$ axis of the theoretical calculations with those of the experiment. 
This is done by aligning the direction of maximum magnetization in the $xy$ plane between simulation and experiment.

The results of the simulations of the dynamics are shown in Fig.~\ref{fig:fig1}(c) for the magnetization and 
Fig.~\ref{fig:SM_data_VS_TDVP} for the correlations. They are in very good agreement with the data, except for
the AFM $x$-correlations, which grow slightly faster than observed. 

\subsection{Ground and thermal state calculations}\label{SubSM:Ground_state_calculations}

We simulate ground states and thermal states of the Hamiltonian $H_{\rm tot}$ using 
the quantum Monte Carlo (QMC) method based on the 
Stochastic Series Expansion approach \cite{Syljuasen_2002} for the FM. 
For the AFM, we use the density matrix renormalization group (DMRG) 
provided by the ITensor package~\cite{Fishman_2022}
with  truncation error $< 10^{-10}$ and bond dimension up to 200 (QMC has a sign problem in that case). 
For both the FM and AFM, we average over a few hundred realizations of holes, 
with a concentration $p$ which accounts both for the holes in the initial state due to the 
imperfect excitation of atoms to Rydberg states and the ones appearing during the ramp. 

In the FM case, both the experimental and numerical data for the ramp dynamics show a 
significant residual magnetization in the $xy$ plane $m_{\rm XY} \approx 0.18$, 
resulting from an imperfect preparation of the initial state of the sequence. 
We include this element in the finite-temperature calculations 
by adding a small transverse field $h/J \approx 10^{-2}$, adjusted so as to stabilize the experimentally observed magnetization.  
For the AFM chains, we use the purification method applied to MPS~\cite{Verstraete2004}. 
The final residual magnetization in the $xy$ plane is very small in the AFM case, 
and there we do not include any transverse field. 

In an ideal state-preparation sequence, the system would be prepared in a zero-magnetization 
state at the beginning of the ramp, and it would remain in that sector throughout the evolution, 
since the Hamiltonian conserves magnetization. As a consequence, the variance of the magnetization 
${\rm Var}(M_z) = \sum_{ij} C^z(i,j)$ would be strictly zero in the final state 
-- at variance with a thermal state with a fluctuating magnetization. 
This constraint is global and is important for observables sensitive to 
scales comparable to the system size, such as the long-distance $C^z(i,j)$ correlations.
In practice, however, the initial state in the experiment has a finite uncertainty on $M^z$ due to the 
finite fidelity of the initial state preparation. 
This variance is then propagated throughout the ramp sequence, although it remains 
significantly smaller than that of thermal states at the temperatures relevant for the experiment. 
In principle one would need to constrain the simulated thermal states to reproduce the 
experimentally observed ${\rm Var}(M_z)_{\rm exp}$, 
adding therefore an additional Lagrange multiplier to the Gibbs ensemble beside temperature 
(and transverse field for the FM). We opt for a simpler approach, 
by leaving the simulations unconstrained, and correcting the simulated correlations $C^z(i,j)$ 
by an offset $C_{\rm off}$, $\tilde C^z(i,j) = C^z(i,j) + C_{\rm off}$,  
such that $\sum_{ij} \tilde C^z(i,j) = {\rm Var}(M_z)_{\rm exp}$. 
This correction affects primarily the long-distance correlations. 
Even though empirical, we verified its quantitative validity using exact diagonalization in small systems. 

As for the temperature, it could be determined in principle by matching the experimental energy 
to that of the numerical simulations. Yet such an approach is highly sensitive to uncertainties in the 
experimental reconstruction of correlations at short range, which dominate the energy. 
We use a more global approach, searching for the temperature which gives the 
best match between experiments and simulations for the whole spatial 
structure of the $C^x$ and $C^z$ correlations.  This leads to the temperature cited in the main text.

\subsection{Simulation of the quench experiment}\label{SubSM:Simulation_of_the_quench_experiment}

The simulations were done using the software ITensor~\cite{Fishman_2022}. 
We used the TDVP algorithm with time step 0.02\,$\mu$s (i.e. $0.0051/J$), 
truncation error~$<10^{-5}$ and bond dimension~1600. They do not include experimental imperfections.

\section{Luttinger-liquid properties vs. localization of dipolar chains}\label{SM:Luttinger}

We analyze here the ground-state correlations of dipolar chains to compare them 
to the expectations from Luttinger-liquid theory, 
in the ideal case and in the presence of holes.  

\begin{figure}
	\centering
	\includegraphics[width=\linewidth]{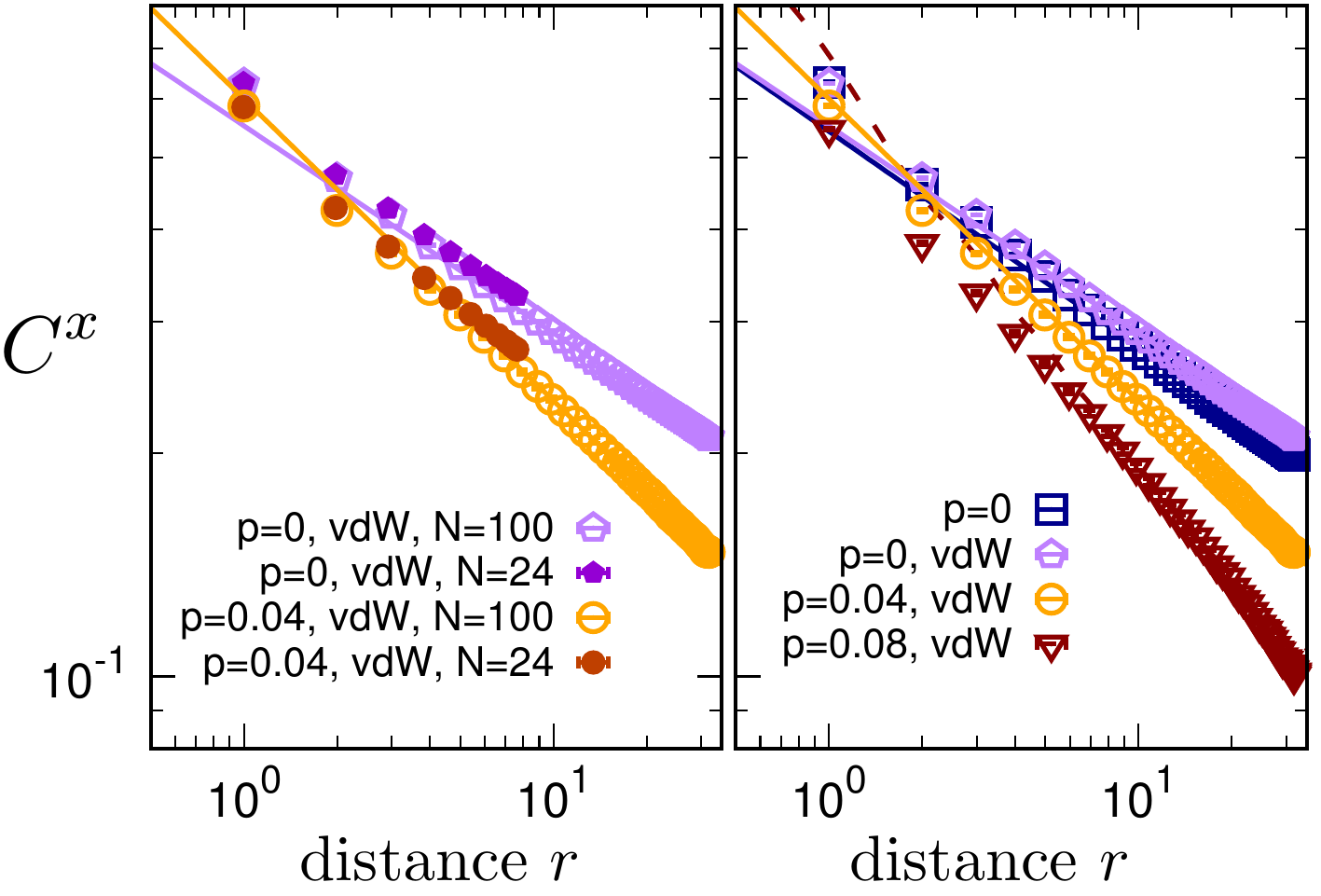}
	\caption{\textbf{Ground-state correlations of the dipolar FM chain.}
	(a) Correlations $Cx$ for the ground state of the dipolar FM Hamiltonian (including vdW interactions) 
	for $p=0$ and  $p=0.04$ hole doping. We compare the experimental system size, 
	$N=24$ to a larger system size, $N=100$. 
	(b) Same as in (a) for $N=100$, including the case of a chain without holes and vdW interactions, 
	and a chain with doping $p=0.08$. 
	In both figures $r$ is the chord distance. 
	The solid/dashed lines are power-law fits to $A/r^{1/(2K)}$. }
	\label{fig:SM_corr_FM}
\end{figure}

\begin{figure}[h!]
	\centering
	\includegraphics[width=\linewidth]{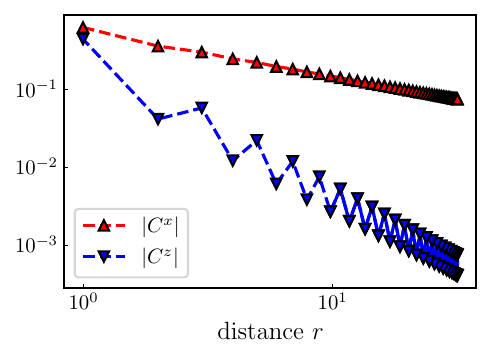}
	\caption{\textbf{Ground-state correlations of the dipolar AFM chain.}
	The dashed lines connect the points, while the solid lines are fits of the long-distance 
	behavior to the expected power-law decay in a LL.}
	\label{fig:SM_corr_AFM}
\end{figure}

\subsection{Dipolar chains without holes}

\noindent{\bf Correlation functions and compressibility.} 
Fig.~\ref{fig:SM_corr_FM} shows QMC results for the ground state of the dipolar FM chain. 
We compare two system sizes, $N=24$ (the experimental one) 
and a larger one ($N=100$) to capture the asymptotic properties of the correlations. 
The calculations were performed at temperatures $T \lesssim 10^{-2} J$ where thermal 
effects are negligible. We observe that already for $N=24$ the correlations decay as $1/r^{1/(2K)}$ --
and the value of $K$ does not change significantly when 
fitting by the more involved Eq.~(2) of the main text.
The fit leads to a value of $K_{\rm FM} = 1.85(1)$, which is larger than the better estimates for the longer chain. 
Indeed, for $N=100$, we obtain 
$K_{\rm FM} \approx 1.72$ for the chain with purely dipolar interactions, 
in agreement with the estimate of \cite{Gupta_2023}, and  $K_{\rm FM, vdW} \approx 1.79$ 
when including the vdW interactions. The increase of $K$ in the presence of ferromagnetic 
vdW interactions is consistent with what is observed in the XXZ model, in which the $K$ 
parameter is a monotonically increasing function of the (ferromagnetic) 
interactions for the $z$ spin components \cite{Giamarchi_2004}. 

From the magnetization curve of the $N=100$ system at small fields (not shown) 
we calculate a spin susceptibility along the $z$ axis of $\kappa \approx 0.33$, 
from which we estimate the sound velocity using $u_{\rm FM}/(2Ja) = \pi \kappa K$ \cite{Giamarchi_2004}. 
We obtain $u/(2Ja) \approx  1.85$ for the system with vdW interactions.  
This represents a large increase with respect to the case 
of nearest neighbor interactions ($K=1$). Thus, compared to the case of NN interactions 
the ferromagnetic dipolar ones lead to an enhancement of ground state correlations, 
as well as to an acceleration of the propagation of modes at low wavevector. 

Fig.~\ref{fig:SM_corr_AFM} shows instead the DMRG results for the correlations 
$C^x$ and $|C^z|$ of the AFM chain, including antiferromagnetic vdW interactions.
 A fit using Eqs.~(2) and (3) of the main text gives a consistent picture of a Tomonaga-Luttinger liquid with 
 $K_{\rm AF}\approx 0.865$, reduced with respect to nearest-neighbor interactions because of frustration. 
 The analysis of the magnetization curve gives $\kappa \approx 0.33$, 
 leading to a sound velocity $u_{\rm AF}/(2Ja) \approx 0.9$, 
 reduced as well with respect to the nearest-neighbor XY chain. \\

\begin{figure*}
	\centering
	\includegraphics[width=\textwidth]{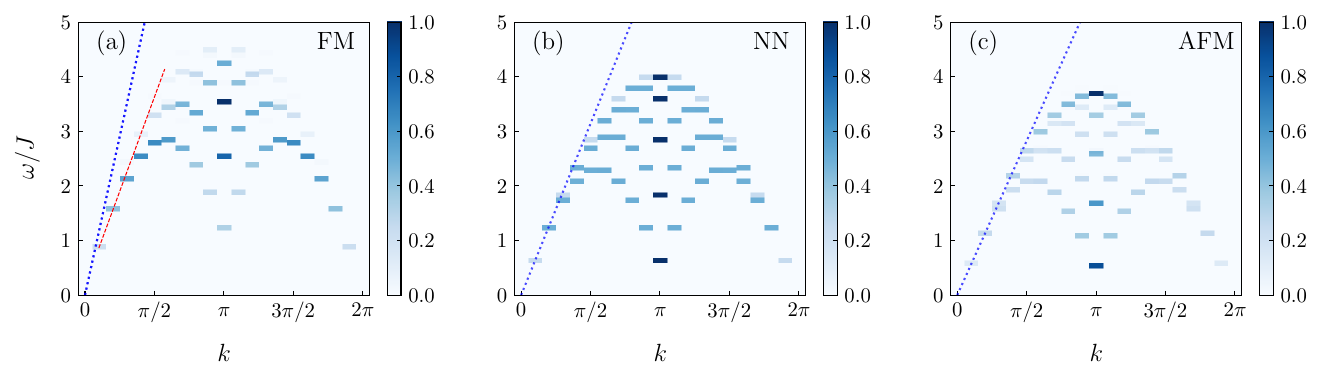}
	\caption{\textbf{Dynamical structure factors of dipolar and NN chains.} 
	(a) Dipolar FM chain; (b) NN XY chain: (c) Dipolar AFM chain. 
	All results have been obtained for a $N=20$ chain via exact diagonalization. 
	The color bar is normalized to the peak value. 
	Blue dashed lines: dispersion of the LL sound mode $\omega = uk$, 
	with $u$ estimated from the ground-state calculations. 
	In panel (a), the dashed red line indicates the effective sound mode $\omega = v_g k$ 
	observed in the experiment, with $v_g = 2.2 Ja$.}
	\label{fig:DSF}
\end{figure*}

\noindent{\bf Dynamical structure factor and sound mode.}
The excitation spectrum of a Tomonaga-Luttinger liquid is  composed of a sound 
mode with dispersion relation $\omega = u k$ \cite{Giamarchi_2004}. 
In the case of a lattice spin system, the sound mode captures only the long-wavelength excitations, 
while at shorter wavelengths the spectrum exhibits a continuum of excitations for each wave vector, 
as revealed e.g. by neutron scattering experiments on quantum magnets \cite{Lake_2005}. 
We test the importance of the sound mode with respect to the other modes of the spectrum 
of the lattice model by calculating the dynamical structure factor in the ground state: 
\begin{equation}
S(q,\omega) \sim \sum_{n} |\langle \psi_n | S^z_q | 0 \rangle|^2  \delta (\omega - \omega_{n0})
\end{equation}  
where $|\psi_0\rangle, |\psi_n\rangle$ are the ground state and a 
generic excited state of the Hamiltonian, respectively, with corresponding energies 
$E_0, E_n$; $S^z_q = \frac{1}{\sqrt{N}} \sum_j e^{iqr_j} \sigma_j^z$ 
and $\omega_n =  (E_n - E_0)/\hbar$. 
Figure~\ref{fig:DSF} shows the dynamical structure factor from exact diagonalization 
(using the QuSpin package \cite{QuSpin}) for the dipolar FM and AFM chains with $N=20$ sites, 
as well as for NN interactions. 
For the FM (Fig.~\ref{fig:DSF}a) 
we observe that the sound mode contains essentially  one wavevector, 
due to the significant curvature of the dispersion relation. 
We understand this curvature from the fact that, compared to the system with NN interactions, 
the sound velocity is renormalized by the dipolar interactions much more 
strongly than the bandwidth of the excitations. 
Compared to the NN XY chain (Fig.~\ref{fig:DSF}b), for which $u/(2Ja) = 1$ and the maximum energy
$\omega_{\rm max} = 4J$, the dipolar FM chain has $u/(2Ja) \approx 1.85$ 
while $\omega_{\rm max} \lesssim 5 J$, i.e. a $85\%$ increase in the 
sound velocity for a $\approx 20\%$ increase in the bandwidth. 
The latter can be understood simply as an effect of the integral of the 
dipolar interactions in 1D, $\sum_{r>0} 1/r^3 \approx 1.2 $. 
The sound velocity is much more renormalized, 
since the model is close to a quantum phase transition to long-range order, 
occurring for interactions $1/r^\alpha$ with $\alpha \lesssim 2.8$, 
at which both $u$ and $K$ diverge \cite{Maghrebi_2017}.
  
In the case of the dipolar AFM chain (Fig.~\ref{fig:DSF}c), both the bandwidth and the
sound velocity are reduced by $\approx 10\%$, 
compatible with the fact that $\sum_{r>0} (-1)^{r+1}/r^3 \approx 0.9 $. 
As a consequence, the sound mode is as pronounced in the spectrum 
as in the case of the NN interactions (Fig.~\ref{fig:DSF}b). 
The Luttinger-liquid sound mode is thus visible in the quench dynamics of 
the correlations in the AFM case, as shown in the main text. 
For the dipolar FM, the sound mode with the highest group velocity 
is not clearly visible on small systems, as it is masked by modes at intermediate 
$k$ with a weaker group velocity. 
Figure~\ref{fig:DSF}(a) shows that the observed light-cone velocity in the 
FM experiment, $u/(2Ja) \approx 1.2$, is compatible with the group velocity 
of the dispersion relation at wavevectors just above the one(s) associated with the actual sound mode. 
  
\subsection{Dipolar chains with holes}\label{subSM:chains_with_holes}

As discussed in the main text, the imperfect preparation of the initial state 
and the decay of atoms during dynamics leads to holes appearing in the chain. 
They break the chain into disconnected segments for NN interactions, 
destroying the power-law correlations of the undoped system. 
Power-law interactions allow instead for a finite concentration of holes, 
while retaining a full connectivity, and possibly preserving the 
Luttinger-liquid nature of the ground state at long distances. \\

\noindent{\bf Hole doping in the FM chain.} We probe the effect of a finite hole 
concentration for the FM via QMC calculations, averaged over $> 100$ hole realizations, 
and at temperatures $T \lesssim 10^{-2} J$ to eliminate thermal effects. 
Figure~\ref{fig:SM_corr_FM}(b) shows the effect of two hole concentrations, 
$p=4\%$ (the one at the end of the  ramp for the FM) 
and $p=8\%$. The $C^x$ correlations with 4\% of holes are consistent 
with those of a Tomonaga-Luttinger liquid with exponent $K_{\rm FM, vdW}(p=0.04) \approx 1.22$, 
while the analysis of the  $C^z$ correlations (not shown) gives $K_{\rm FM, vdW}(p=0.04) \approx 1.28(3)$. 
This suggests that 4\% of hole doping preserves the LL physics in the FM chain, 
albeit renormalizing significantly the Luttinger parameters. 
To compare to the experiment, we  calculate $K$ on an $N=24$ ring, 
thus providing an upper bound to the actual $K$ value (see Fig.~\ref{fig:SM_corr_FM}a). 
We find that for $p=4\%$, $K_{\rm FM}(N=24) = 1.55(1)$ from the $C^x$ correlations 
and $K_{\rm FM}(N=24) = 1.44(1)$ from the $C^z$ correlations. 

The simulation for $p=8\%$ leads  instead to $K_{\rm FM, vdW}(p=0.08) \approx 0.93$ 
from the $C^x$ correlations and  $K_{\rm FM, vdW}(p=0.08) \approx 1.20(4)$ from $C^z$ correlations. 
The fact that one value lies below 1 and that the $C^x$ correlations deviate from a strict power-law behavior 
suggests that the chain with doping sits in the vicinity of a quantum phase transition from 
Luttinger-liquid physics to a different, disorder-induced phase. 
This new phase phase could be a localized one such as a Bose glass \cite{Giamarchi_1988}, 
or a Mott glass \cite{Altman_2004}.
The latter may be favored by the fact that hole doping leads to bond disorder 
for the remaining spins, which is accompanied by a net zero magnetization, 
i.e. a commensurate (one-half) filling in the bosonic language. 
A third possibility is that the bond-disordered chain realizes a random-singlet phase, 
which retains power-law correlations \cite{Fisher_1994}. 

In fact, we cannot exclude that even the smaller dilution we considered, $p = 4\%$, 
leads to a stabilization of a random-singlet phase in very large systems. 
The fact that we observe an effective LL behavior on the system sizes we simulated 
may be due to a crossover from a LL-like behavior at short distance 
to a random-singlet behavior at long distance, 
for which all two-point correlations are expected to decay as $1/r^2$ \cite{Fisher_1994}. 
This crossover, observed in microscopic calculations on bond-disordered chains 
\cite{Laflorencie_2004,Hoyos_2007}, is expected to occur over length scales which 
diverge with the inverse strength of disorder, controlled in our case by the density of holes. \\

\begin{figure}
	\centering
	\includegraphics[width=0.9\linewidth]{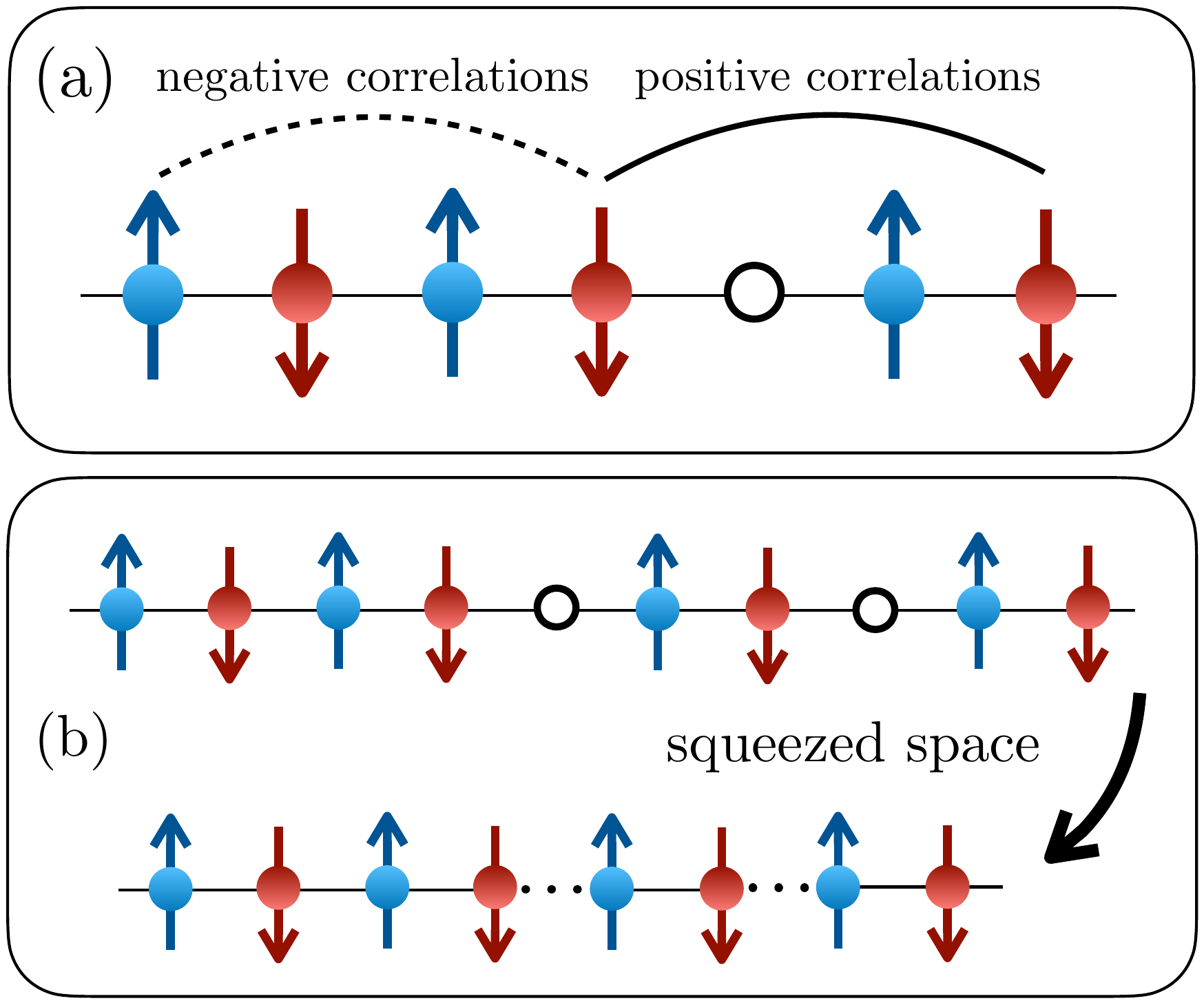}
	\caption{\textbf{Hole doping in the dipolar AFM}. 
	(a) Hole doping leads to a slip of the sublattice structure, such that spins at the same distance 
	can be correlated or anticorrelated, depending on the even or odd number of holes between them; 
	(b) The sublattice slip can be removed by going to ``squeezed space", i.e. eliminating the holes 
	(provided that one knows their positions). Nonetheless this leaves behind some randomness in the couplings: 
	e.g. the NN couplings -- indicated in the lower part of the figure -- are bimodally distributed 
	(if one neglects two or more adjacent holes), with values $J$ (solid lines) and $J/8$ dotted lines. }
	\label{fig:holes_AFM}
\end{figure}

\begin{figure}
	\centering
	\includegraphics[width=0.9\linewidth]{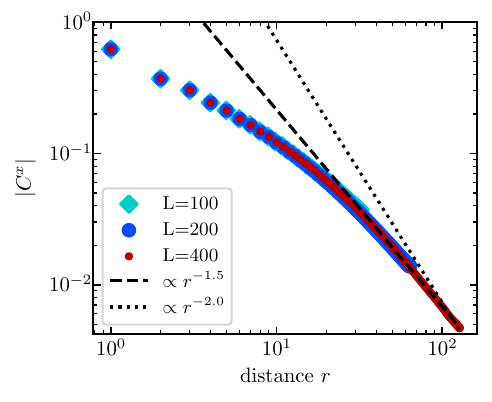}
	\caption{\textbf{Correlations in a bond-disordered NN XY chain.} 
	The calculation was done on chains with $N=100, 200$ and 400 spins. 
	The solid lines correspond to a power-law fit of the correlation tail, 
	as well as the expected $r^{-2}$ decay in the random-singlet phase. }
	\label{fig:Cx_disnnXY}
\end{figure}

\noindent{\bf Hole doping in the AFM chain.}
Contrary to the FM chain, hole doping in the AFM is expected to 
affect strongly LL physics. 
First, hole doping leads to a disruption of the sublattice structure in a 1D AFM 
with power-law interactions: every hole leads to a one-site slip of the sublattice structure, 
so that correlations between two sites at the same distance can be of opposite signs 
if a hole appears or not in between the two sites (see Fig.~\ref{fig:holes_AFM}a). 
This effect is the one mainly responsible for the observation, shown in Fig.~2(f) of the main text, 
that even in the ground states the $C^x$ correlations of the hole-doped AFM 
exhibit a faster-than-algebraic decay. 
Yet this sublattice slip does not affect the correlations at long distance, 
provided that one knows the positions of the holes: 
we could define correlations in ``squeezed space" 
(Fig~\ref{fig:holes_AFM}b) \cite{Hilker_2017}, eliminating the holes from the picture. 
These correlations may  retain a power-law behavior 
-- this would be  true if the spin couplings in squeezed space 
were the same as the couplings in a clean system. In the case of dipolar coupling, 
the interactions are reduced by a factor of 1/8 across a hole with 
respect to those between nearest neighbors. As a consequence, 
the spin model that emerges in squeezed space is an XY model with bond disorder. 
 
We expect this bond disorder to strongly affect LL physics in the AFM chain for the hole concentrations 
$p$ realized in the experiment. We motivate this conclusion by considering that, in the case of NN interactions, 
the model can be mapped onto free fermions \cite{Lieb_1961}, which in 
1D are strongly susceptible to disorder. A disordered potential would lead to localization 
of the entire spectrum. Disordered hoppings with a broad distribution 
leads instead to a random-singlet phase \cite{Fisher_1994}, which, unlike localized phases, retains 
power-law correlations.
Hole doping leads to bond disorder with  a 
discrete (nearly bimodal) distribution, hence the relevance of the random-singlet 
phase is not immediately obvious.  
We hence probe numerically the correlations 
of the NN XY chain with bimodal bond disorder, with  $p=6\%$ of 
bonds with couplings $J/8$ randomly doped in the system, the other bonds having couplings $J$.
Figure~\ref{fig:Cx_disnnXY} shows the $C^x$ correlations for this system, obtained 
by a mapping onto free fermions. We observe correlations which are 
incompatible with a power-law decay, although an emergent power-law decay  
seems to appear at very large distances. This behavior is suggestive of a 
random-singlet phase, although further studies would be needed to 
precisely pinpoint the nature of the ground state.  
 
In the case of the dipolar AFM chain, the undoped ideal system has $K<1$, 
leading to weaker $C^x$ correlations than the NN XY chain. Hence we expect 
that a fraction $p = 6\%$ of holes, as in the experiment, 
will lead {\it a fortiori} to a disorder-induced phase. 
We therefore conclude that the ground state of the hole-doped dipolar AFM 
is not a Tomonaga-Luttinger liquid for the hole concentrations relevant to the experiment, 
but rather a disorder-dominated phase. 

\section{Friedel oscillations}\label{SM:Friedel_oscillations}

\subsection{Analytic origin of the Friedel oscillation}\label{SM:Analytic_derivation_of_the_Friedel_oscillation}

The open-chain ground states, at each filling fraction, can be understood as 
those of a conformal field theory (CFT) with open boundary conditions---a so-called boundary 
CFT~\cite{cardyConformalInvarianceSurface1984,burkhardtUniversalOrderparameterProfiles1985}.
In this general setting, primary field excitations $\mathcal{O}$ of the CFT will develop 
an expectation value $\langle \mathcal{O} \rangle$ that decays away from the boundary 
as $r^{-\Delta_{\mathcal{O}}}$, where $\Delta_{\mathcal{O}}$ is that field's scaling dimension.
(Under periodic boundary conditions, $\langle \mathcal{O}\rangle=0$.)
For the 1+1d case at hand, the explicit functional form can be calculated via a conformal 
mapping to be
$\langle \mathcal{O} \rangle = A_{\mathcal{O}} \left[\frac{L}{\pi}\cos\left(\frac{\pi j}{L}\right) \right]^{-\Delta_\mathcal{O}}$~\cite{fathLuttingerLiquidBehavior2003}.
Here, $j \in \left\llbracket-\frac{N-1}{2},\frac{N-1}{2}\right\rrbracket$ is the position of the spins (in units of sites), 
with $j=0$ being the center of the chain.
This equation holds when the long-wavelength scaling limit is meaningful, 
i.e. at some sufficiently large distance from the boundary.
The coefficient $A_{\mathcal{O}}$ is generically nonzero for all operators 
not forbidden by any remaining microscopic symmetry.
We note that in some cases the critical exponents in the boundary CFT 
can differ from those with PBC, but in our instance they are the same.
The experimental measurements of $\langle \sigma^z_j \rangle$ probe two primary 
operators of the LL CFT: the conserved current $J = \partial_x \phi$ and the 
vertex operator $\mathcal{V}_{1,0} = e^{i\phi}$~\cite{Ginsparg:1988ui}. 
The first corresponds to the zero-momentum component of $\sigma^z$, i.e. the average magnetization 
$m_0$; it has dimension $\Delta_J=1$, and is responsible for the $r^{-2}$ 
part of the $C^z$ decay in the periodic system.
The second has scaling dimension $\Delta_{1,0}=K$, and corresponds to oscillations of 
$\sigma^z$ at a particular momentum which depends on $m_0$.
As noted in the main text, consistency requirements between the actions of microscopic and 
continuum-limit symmetries set this wavevector to be 
$\pi(1/2-m_0)$~\cite{elseNonFermiLiquidsErsatz2021, chengLiebSchultzMattisLuttingerHooft2023}.
We therefore expect that,
\begin{multline}
	\langle \sigma_j^z \rangle - m_0 =  \langle \mathcal{V}_{1,0} \rangle + \cdots
	\\ =  A \cos(2 k_F j + \delta)\left[\frac{L}{\pi} \cos\left(\frac{\pi j}{L}\right)\right]^{-K} + \cdots,
	\label{eq:Friedel_CT}
\end{multline}
where the ellipses denote more rapidly-decaying terms, $A$ is a non--universal prefactor, 
and $\delta$ is the overall phase for the oscillations.
For our odd-$N$ chain, reflection symmetry across the center site ($j=0$) 
sets $\delta=0$, leading to Eq.~(\ref{eq:Friedel}) in the main text.

\subsection{Background subtraction for Friedel oscillations}\label{SM:Background_subtraction_for_Friedel_oscillations}

\begin{figure*}
	\centering
	\includegraphics{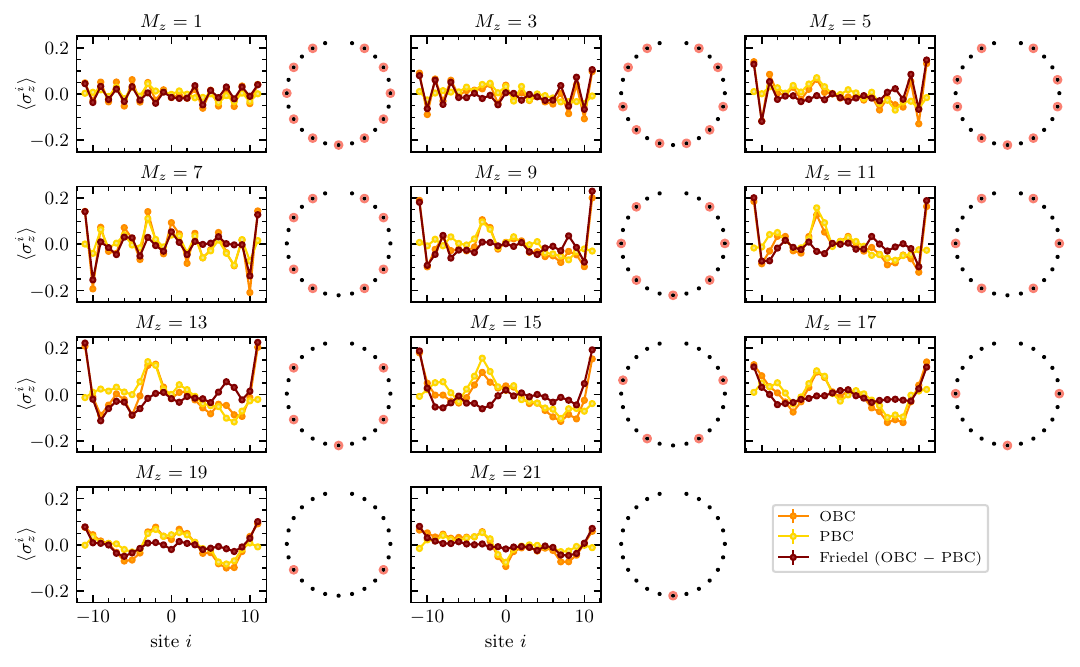}
	\caption{\textbf{Extraction of the magnetization for Friedel oscillations.}
	Spatial dependence of the $z$-magnetization $\langle \sigma_z^i \rangle$ 
	with the site positions $i$ (the average magnetization 
	$\sum_{j=1}^N \langle \sigma_z^j \rangle \; / \; N$ is removed for clarity), 
	for different $M_z$ sectors. For each value $m$ of the total magnetization, 
	a different initial state $\ket{\psi_m}$ is used for the adiabatic preparation, 
	with $(N+m)/2$ spins in $\ket{\uparrow}$ (non-addressed atoms) and 
	$(N-m)/2$ spins in $\ket{\downarrow}$ (addressed atoms); 
	the position of the addressed atoms is indicated by pink circles.
	The Friedel signal (OBC $-$ PBC) at site $i$ is the difference of the OBC 
	magnetization and the PBC magnetization measured with the same 
	sequence and the same addressing pattern. 
	The ramp times differ for the different $M_z$ sectors 
	($T=2500$~ns for $M_z=1$, $T=5000$~ns for 
	$M_z\in[3,5,7,9,11,13,15]$, $T=8000$~ns for $M_z\in[17,19,21]$), 
	and the measurements were always taken $500$~ns after the end of the ramp.
	}
	\label{fig:SM_Friedel}
\end{figure*}

The ideal ground state of a closed ring (PBC) does not break translational invariance, 
and in particular we expect no Friedel oscillations. To check this experimentally 
we use the same adiabatic sequence as in the open ring system. 
The resulting magnetization $\langle \sigma_z^i \rangle_{\rm PBC}$ 
is shown in Fig.~\ref{fig:SM_Friedel} (yellow points), for different addressing 
pattern corresponding to all target values of the total magnetization $M_z$. 
In contrast with the expectation for the ideal ground state, we observe a spatially inhomogeneous 
magnetization which depends on the addressing pattern. 
We also distinguish a small gradient of magnetization 
(which in a circular geometry translates into an oscillation with wavelength $N$). 
The amplitude of those inhomogeneities are on the same order 
as the expected Friedel signal in a chain with OBC ($\langle \sigma_z^i \rangle \sim 0.1$). 
Several effects could explain the inhomogeneities: non-adiabaticity of the ramp 
leading to a reminiscence of the initial addressing pattern; 
positional disorder on the average atomic position; 
inhomogeneous spin frequencies due to gradients of electric or magnetic field.

When we perform the same experiment on the open ring, 
the magnetization $\langle \sigma_z^i \rangle_{\rm OBC}$ shows very similar inhomogeneities, 
on top of well-defined oscillations close to the edges [orange points of Fig.~\ref{fig:SM_Friedel}]. 
To mitigate the measured inhomogeneities and conserve only the contribution of the edges, 
we subtract the PBC background to the Friedel OBC signal: 
$\langle \sigma_z^i \rangle_{\rm Friedel} = \langle \sigma_z^i \rangle_{\rm OBC} - 
\langle \sigma_z^i \rangle_{\rm PBC}$ [red points of Fig.~\ref{fig:SM_Friedel}]. 
The data shown in Fig.~\ref{fig:fig3}(c) of the main text 
corresponds to the corrected data $\langle \sigma_z^i \rangle_{\rm Friedel}$.

\bibliography{References}

\end{document}